\documentclass[universe,review,submit,moreauthors,pdftex,10pt,a4paper]{mdpi} 
\firstpage{1} 
\articlenumber{x}
\doinum{10.3390/------}
\pubvolume{xx}
\pubyear{2016}
\copyrightyear{2016}
\history{Received: date; Accepted: date; Published: date}

 \theoremstyle{mdpi}
 \newcounter{thm}
 \newcounter{ex}
 \newcounter{re}
 \theoremstyle{mdpidefinition}

\usepackage{aas_macros}
\usepackage{bm}
\newcommand{\bs}{\bar s}
\usepackage{rotating}
\usepackage{placeins} 
\Title{Tests of Lorentz symmetry in the gravitational sector}

\Author{Aur\'elien Hees $^{1,}$*, Quentin G. Bailey $^2$, Adrien Bourgoin $^{3}$, H\'el\`ene Pihan-Le Bars $^{3}$, Christine Guerlin $^{4,3}$ and Christophe Le Poncin-Lafitte $^3$}
\AuthorNames{Aurelien Hees, Quentin G. Bailey, Adrien Bourgoin, Helene Pihan-Le Bars, Christine Guerlin and  Christophe Le Poncin-Lafitte}

\address{%
$^{1}$ \quad Department of Physics and Astronomy, University of California, Los Angeles, CA 90095, USA\\
$^{2}$ \quad Physics Department, Embry-Riddle Aeronautical University, Prescott, AZ 86301, USA \\
$^{3}$ \quad SYRTE, Observatoire de Paris, PSL Research University, CNRS, Sorbonne Universit\'es, UPMC Univ. Paris 06, LNE, 61 avenue de l'Observatoire, 75014 Paris, France \\
$^{4}$ \quad Laboratoire Kastler Brossel, ENS-PSL Research University, CNRS, UPMC-Sorbonne Universit\'es, Coll\`ege de France, 75005 Paris, France}

\corres{Correspondence: ahees@astro.ucla.edu; Tel.: +1-310-825-8345}

\abstract{Lorentz symmetry is one of the pillars of both General Relativity and the Standard Model of particle physics. Motivated by ideas about quantum gravity,  unification theories and  violations of  CPT symmetry, a significant effort has been put the last decades into testing Lorentz symmetry. This review focuses on Lorentz symmetry tests performed in the gravitational sector. We briefly review the basics of the pure gravitational sector of the Standard-Model Extension (SME) framework, a formalism developed in order to systematically parametrize hypothetical violations of the Lorentz invariance. Furthermore, we discuss the latest constraints obtained within this formalism including analyses of the following measurements: atomic gravimetry, Lunar Laser Ranging, Very Long Baseline Interferometry, planetary ephemerides, Gravity Probe B, binary pulsars, high energy cosmic rays, \dots \ In addition, we propose a combined analysis of all these results. We also discuss possible improvements on current analyses and present some sensitivity analyses for future observations.}

\keyword{Experimental tests of gravitational theories; Lorentz and Poincar\'e invariance; Modified theories of gravity; Celestial mechanics; Atom interferometry; Binary pulsars.}

\begin{document}


\section{Introduction}
The year 2015 was the centenary of the theory of General Relativity (GR), the current paradigm for describing the gravitational interaction (see e.g. the Editorial of this special issue \cite{iorio:2015kx}). Since its creation, this theory has passed all experimental tests with flying colors \cite{will:2014la,turyshev:2009qv} ; the last recent success was the discovery of gravitational waves \cite{abbott:2016ys}, summarized in \cite{cervantes-cota:2016ly}. On the other hand, the three other fundamental interactions of Nature are described within the Standard Model of particle physics, a framework based on relativistic quantum field theory. Although very successful so far, it is commonly admitted that these two theories are not the ultimate description of Nature but rather some effective theories. This assumption is motivated by the construction of a quantum theory of gravitation that has not been successfully developed so far and by the development of a theory that would unify all the fundamental interactions. Moreover, observations requiring the introduction of Dark Matter and Dark Energy also challenge GR and the Standard Model of particle physics since they cannot be explained by these two paradigms altogether \cite{debono:2016zr}. It is therefore extremely important to test our current description of the four fundamental interactions \cite{berti:2015ve}.

Lorentz invariance is one of the fundamental symmetry of relativity, one of the corner stones of both GR and the Standard Model of particle physics. It states that the outcome of any local experiment is independent of the velocity and of the orientation of the laboratory in which the experiment is performed. If one considers non-gravitational experiments, Lorentz symmetry is part of the Einstein Equivalence Principle (EEP). A breaking of Lorentz symmetry implies that the equations of motion, the particle thresholds, etc... may be different when the experiment is boosted or rotated with respect to a background field \cite{colladay:1997vn}. More precisely, it is related to a violation of the invariance under  ``particle Lorentz transformations''~\cite{colladay:1997vn}  which are the boosts and rotations that relate the properties of two systems within a specific oriented inertial frame (or in other words they are boosts and rotations on localized fields but not on background fields).  On the other hand, the invariance under coordinates transformations known as ``observer Lorentz transformations''~\cite{colladay:1997vn}  which relate observations made in two inertial frames with different orientations and velocities is always preserved.  Considering the broad field of applicability of this symmetry, searches for Lorentz symmetry breaking provide a powerful test of fundamental physics. Moreover, it has been suggested that Lorentz symmetry may not be a fundamental symmetry of Nature and may be broken at some level. While some early motivations came from string theories \cite{kostelecky:1989jk,kostelecky:1989yu,kostelecky:1991nr}, breaking of Lorentz symmetry also appears in loop quantum gravity \cite{gambini:1999aa,amelino-camelia:2013aa,mavromatos:2005aa,myers:2003aa}, non commutative geometry \cite{hayakawa:2000aa,carroll:2001aa}, multiverses \cite{bjorken:2003aa}, brane-world scenarios \cite{burgess:2002aa,frey:2003aa,cline:2004aa} and others (see for example \cite{tasson:2014qv,mattingly:2005uq}).

Tests of Lorentz symmetry have been performed since the time of Einstein but the last decades have seen the number of tests increased significantly \cite{kostelecky:2011ly} in all fields of physics. In particular, a dedicated effective field theory has been developed in order to systematically consider all hypothetical violations of the Lorentz invariance. This framework is known as the Standard-Model Extension (SME) \cite{colladay:1997vn,colladay:1998ys} and covers all fields of physics. It contains the Standard Model of particle physics, GR and all possible Lorentz-violating terms that can be constructed at the level of the Lagrangian, introducing a large numbers of new coefficients that can be constrained experimentally. 

In this review, we focus on the gravitational sector of the SME which parametrizes deviations from GR. GR is built upon two principles \cite{thorne:1971fk,will:1993fk,will:2014la}: (i) the EEP and (ii) the Einstein field equations that derive from the Einstein-Hilbert action. The EEP gives a geometric nature to gravitation allowing this interaction to be described by spacetime curvature. From a theoretical point of view, the EEP implies the existence of a spacetime metric to which all matter minimally couples~\cite{thorne:1973fk}. A modification of the matter part of the action will lead to a breaking of the EEP. In SME, such a breaking of the EEP is parametrized (amongst others) by the matter-gravity coupling coefficients $\bar a_\mu$ and $\bar c_{\mu\nu}$ \cite{kostelecky:2011kx,tasson:2016fk}. From a phenomenological point of view, the EEP states that \cite{will:1993fk,will:2014la}: (i) the universality of free fall (also known as the weak equivalence principle) is valid, (ii) the outcome of any local non-gravitational experiment is independent of the velocity of the free-falling reference frame in which it is performed and (iii) the outcome of any local non-gravitational experiment is independent of where and when in the universe it is performed. The second part of Einstein theory concerns the purely gravitational part of the action (the Einstein-Hilbert action) which is modified in SME to introduce hypothetical Lorentz violations in the gravitational sector. This review focuses exclusively on this kind of Lorentz violations and not on breaking of the EEP.

A lot of tests of GR have been performed in the last decades (see \cite{will:2014la} for a review). These tests rely mainly on two formalisms: the parametrized post-Newtonian (PPN) framework and the fifth force formalism. In the former one, the weak gravitational field spacetime metric is  parametrized by 10 dimensionless coefficients \cite{will:1993fk} that encode deviations from GR. This formalism therefore provides a nice interface between theory and experiments. The PPN parameters have been constrained by a lot of different observations \cite{will:2014la} confirming the validity of GR. In particular, three PPN parameters encode violations of the Lorentz symmetry: the $\alpha_{1,2,3}$ PPN coefficients. In the fifth force formalism, one is looking for a deviation from Newtonian gravity where the gravitational potential takes the form of a Yukawa potential characterized by a length $\lambda$ and a strength $\alpha$ of interaction \cite{fischbach:1986uq,talmadge:1988uq,fischbach:1999ly,adelberger:2003uq}. These two parameters are very well constrained as well except at very small and large distances (see \cite{reynaud:2005ys}). 

The gravitational sector of SME offers a new framework to test GR by parametrizing deviations from GR at the level of the action, introducing new terms that are breaking Lorentz symmetry. The idea is to extend the standard Einstein-Hilbert action by including Lorentz-violating terms constructed by contracting new fields with some operators built from curvature tensors and covariant derivatives with increasing mass dimension~\cite{bailey:2016aa}. The lower mass dimension (dimension 4) term is known as the minimal SME and its related new fields can be split into a scalar part $u$, a symmetric trace free part $s^{\mu\nu}$ and a traceless piece $t^{\kappa\lambda\mu\nu}$.  In order to avoid conflicts with the underlying Riemann geometry, the Lorentz violating coefficients can be assumed to be dynamical fields and the Lorentz violation to arise from a spontaneous symmetry breaking \cite{kostelecky:2004fk,bluhm:2005gf,bailey:2006uq,bluhm:2007ys,bluhm:2008vn,bluhm:2015kx}. The Lorentz violating fields therefore acquire a non-vanishing vacuum expectation value (denoted by a bar). It has been shown that in the linearized gravity limit the fluctuations around the vacuum values can be integrated out so that only the vacuum expectation values of the SME coefficients influence observations~\cite{bailey:2006uq}. In the minimal SME, the coefficient $\bar u$ corresponds to a rescaling of the gravitational constant and is therefore unobservable and the coefficients $\bar t^{\kappa\lambda\mu\nu}$ do not play any role at the post-Newtonian level, a surprising phenomenon known as the t-puzzle~\cite{bailey:2015fk,bonder:2015aa}. The $\bar s^{\mu\nu}$ coefficients lead to modifications from GR that have thoroughly been investigated in \cite{bailey:2006uq}. In particular, the SME framework extends standard frameworks such as the PPN or fifth force formalisms meaning that ``standard'' tests of GR cannot  directly  be translated into this formalism.

In the last decade, several measurements have been analyzed within the gravitational sector of the minimal SME framework: Lunar Laser Ranging (LLR) analysis \cite{battat:2007uq,bourgoin:2016yu}, atom interferometry \cite{muller:2008kx,chung:2009uq}, planetary ephemerides analysis \cite{iorio:2012zr,hees:2015sf}, short-range gravity \cite{bennett:2011vn}, Gravity Probe B (GPB) analysis \cite{bailey:2013kq}, binary pulsars timing \cite{shao:2014rc,shao:2014qd}, Very Long Baseline Interferometry (VLBI) analysis \cite{le-poncin-lafitte:2016yq} and \v Cerenkov radiation \cite{kostelecky:2015db}. In addition to the minimal SME, there exist some higher order Lorentz-violating curvature couplings in the gravity sector \cite{bailey:2015fk} that are constrained by short-range experiments \cite{shao:2015uq,long:2015kx,shao:2016aa}, \v Cerenkov radiation \cite{kostelecky:2015db,tasson:2016fk} and gravitational waves analysis \cite{kostelecky:2016nx,yunes:2016zr}. Finally, some SME experiments have been used to derive bounds on spacetime torsion~\cite{kostelecky:2008aa,heckel:2008aa}. A review for these measurements can be found in \cite{tasson:2016fk}. The classic idea to search for or to constrain Lorentz violations in the gravitational sector is to search for orientation or boost dependence of an observation. Typically, one will take advantage of modulations that will occur through an orientation dependence of the observations due to the Earth's rotation, the motion of satellites around Earth (the Moon or artificial satellites), the motion of the Earth (or other planets) around the Sun, the motion of binary pulsars, \dots The main goal of this communication is to review all the current analyses performed in order to constrain Lorentz violation in the pure gravitational sector.

Two distinct procedures have been used to analyze data within the SME framework. The first procedure consists in deriving analytically the signatures produced by the SME coefficients on some observations. Then, the idea is to fit these signatures within residuals obtained by a data analysis performed in pure GR. This approach has the advantage to be relatively easy and fast to perform. Nevertheless, when using this postfit approach, correlations with other parameters fitted in the data reduction are completely neglected and may lead to overoptimistic results. A second way to analyze data consists in introducing the Lorentz violating terms directly in the modeling of observables and in the global data reduction. In this review, we highlight the differences between the two approaches.

In this communication, a brief theoretical review of the SME framework in the gravitational sector is presented in Section~\ref{sec:theory}. The two different approaches to analyze data within the SME framework (postfit analysis versus full modeling of observables within the SME framework) are discussed and compared in Section~\ref{sec:postfit}. Section~\ref{sec:data} is devoted to a discussion of the current measurements analyzed within the SME framework. This discussion includes a general presentation of the measurements, a brief review of the effects of Lorentz violation on each of them, the current analyses performed with real data and a critical discussion. A  ``grand fit'' combining all existing analyses is also presented. In Section~\ref{sec:future}, some future measurements that are expected to improve the current analyses are developed. Finally, our conclusion is presented in Section~\ref{sec:conclusion}.

\section{The Standard-Model Extension in the gravitational sector}\label{sec:theory}
Many of the tests of Lorentz and CPT symmetry have been analyzed within an effective field theory framework which generically describes possible deviations from exact Lorentz and CPT invariance \cite{colladay:1997vn,colladay:1998ys} and contains some traditional test frameworks as limiting cases \cite{kostelecky:2002uq, kostelecky:2009yu}.
This framework is called,  for historical reasons,  the Standard-Model Extension (SME). One part of the activity has been a resurgence of interest in tests of relativity in the Minkowski spacetime context, where global Lorentz symmetry is the key ingredient. Numerous experimental and observational constraints have been obtained on many different types of hypothetical Lorentz and CPT symmetry violations involving matter \cite{kostelecky:2011ly}.
Another part, which has been developed more recently, has seen the SME framework extended to include the curved spacetime regime \cite{kostelecky:2004fk}. Recent work shows that there are many ways in which the spacetime symmetry foundations of GR can be tested \cite{bailey:2006uq,kostelecky:2011kx}.

In the context of effective field theory in curved spacetime,  violations of these types can be described by an action that contains  the usual Einstein-Hilbert term of GR,  a matter action, plus a series of terms describing Lorentz violation for gravity and matter in a generic way. While the fully general coordinate invariant version of this action has been studied in the literature, we focus on a limiting case that is valid for weak-field gravity and can be compactly displayed. Using an expansion of the spacetime metric around flat spacetime, 
$g_{\mu\nu} = \eta_{\mu\nu}+h_{\mu\nu}$,  the effective Lagrange density to quadratic order in $h_{\mu\nu}$ can be written in a compact form as
\begin{equation}
{\cal L} = {\cal L}_{\rm EH} + \frac {c^3}{32 \pi G} h^{\mu\nu} \bs^{\alpha\beta} {\cal G}_{\alpha\mu\nu\beta}+..., 
\label{lagrangian}
\end{equation}
where ${\cal L}_{\rm EH}$ is the standard Einstein-Hilbert term, ${\cal G}_{\alpha\mu\nu\beta}$ is the double dual of the Einstein tensor linearized in $h_{\mu\nu}$, $G$ the bare Newton constant  and $c$ the speed of light in a vacuum. The Lorentz-violating effects in this expression are controlled by the $9$ independent coefficients in the traceless and dimensionless $\bs^{\mu\nu}$ \cite{bailey:2006uq}.
These coefficients are treated as constants in asymptotically flat cartesian coordinates. The ellipses represent additional terms in a series including terms that break CPT symmetry for gravity; such terms are detailed elsewhere \cite{bailey:2015fk, kostelecky:2015db, kostelecky:2016nx} and are part of the so-called nonminimal SME expansion. Note that the process by which one arrives at the effective quadratic Lagrangian (\ref{lagrangian}) is consistent with the assumption of the spontaneous breaking of local Lorentz symmetry, which is discussed below.

Also of interest are the matter-gravity couplings. This form of Lorentz violation can be realized in the classical point-mass limit of the matter sector. In the minimal SME the point-particle action can be written as
\begin{equation}
S_{\rm Matter}=\int d\lambda \, c \left(-m \sqrt{-(g_{\mu \nu}+2c_{\mu \nu})u^\mu u^\nu} -a_\mu u^\mu\right), \label{matter}
\end{equation}
where the particle's worldline tangent is $u^\mu=dx^\mu/d\lambda$ \cite{kostelecky:2011kx}.  The coefficients controlling local Lorentz violation for matter are $c_{\mu \nu}$ and $a_\mu$. In contrast to $\bs^{\mu\nu}$,  these coefficients depend on the type of point mass (particle species) and so they can also violate the EEP. When the coefficients $\bs_{\mu \nu}$, $c_{\mu \nu}$,  and $a_\mu$ vanish perfect local Lorentz symmetry for gravity and matter is restored. It is also interesting to mention that this action with fixed (but not necessarily constant) $a_\mu$ and $c_{\mu\nu}$ represents motion in a Finsler geometry \cite{kostelecky:2011aa,kostelecky:2010aa}.

It has been shown that explicit local Lorentz violation is generically incompatible with Riemann geometry \cite{kostelecky:2004fk}. One natural way around this is assumption of spontaneous Lorentz-symmetry breaking. In this scenario, the tensor fields in the underlying theory acquire vacuum expectation values through a dynamical process.
Much of the literature has been devoted to studying this possibility in the last decades \cite{kostelecky:1989jk,jacobson:2001qf,jackiw:2003mz,bluhm:2005gf,hernaski:2014ul,balakin:2014pd,yagi:2014fk,yagi:2014jk,hernaski:2014lq, seifert:2009rr,kostelecky:2005uq,kostelecky:2009kx,altschul:2010fk}, including some original work on spontaneous Lorentz-symmetry breaking in string field theory \cite{kostelecky:1989yu,kostelecky:1991nr}.
For the matter-gravity couplings in Eq. (\ref{matter}), the coefficient fields  $c_{\mu \nu}$,  and $a_\mu$ are then expanded around their background (or vacuum) values ${\bar c}_{\mu \nu}$,  and ${\bar a}_\mu$. Both a modified spacetime metric $g_{\mu \nu}$ and modified point-particle equations of motion result from the spontaneous breaking of Lorentz symmetry. In the linearized gravity limit these results rely only on the vacuum  values ${\bar c}_{\mu \nu}$,  and ${\bar a}_\mu$. The dominant signals for Lorentz violation controlled by these coefficients are revealed in the calculation of observables in the post-Newtonian limit.

Several novel features of the post-Newtonian limit arise in the SME framework. It was shown in Ref.\ \cite{bailey:2006uq} that a subset of the $\bs^{\mu\nu}$ coefficients can be matched to the PPN formalism \cite{will:1993fk,will:2014la}, but others lie outside it. For example,  a dynamical model of spontaneous Lorentz symmetry breaking can be constructed  from an antisymmetric tensor field $B_{\mu\nu}$ that produces $\bs^{\mu\nu}$ coefficients  that cannot be reduced to an isotropic diagonal form in any coordinate system, thus lying outside the PPN assumptions \cite{altschul:2010fk}. We can therefore see that the SME framework has a partial overlap with the PPN framework,  revealing new directions to explore in analysis via the $\bs^{\mu\nu}$,  ${\bar c}_{\mu \nu}$,  and ${\bar a}_\mu$ coefficients.
The equations of motion for matter are modified by the  matter-gravity coefficients for Lorentz violation ${\bar c}_{\mu \nu}$ and ${\bar a}_\mu$,  which can depend on particle species,  thus implying that these coefficients also control EEP violations. One potentially important class of experiments from the action (\ref{matter}) concerns the Universality of Free Fall of antimatter whose predictions are discussed in \cite{kostelecky:2011kx,kostelecky:2015aa}. In addition,  the post-Newtonian metric itself receives contributions from the matter coefficients ${\bar c}_{\mu \nu}$ and ${\bar a}_\mu$. So for example, two (chargeless) sources with the same total mass but differing composition  will yield gravitational fields of different strength.

For solar-system gravity tests,  the primary effects due to the nine coefficients ${\bar s}^{\mu \nu}$ can be obtained from the post-Newtonian metric and the geodesic equation for test bodies. A variety of ground-based and space-based tests can measure these coefficients \cite{bailey:2009fk,bailey:2010ys,tso:2011uq}. Such tests include Earth-laboratory tests with gravimeters,  lunar and satellite laser ranging, studies of the secular precession of orbital elements in the solar system,  and orbiting gyroscope experiments, and also classic effects such as the time delay and bending of light around the Sun and Jupiter. Furthermore,  some effects described by the Lagrangian (\ref{lagrangian}) can be probed by analyzing data from binary pulsars and measurements of cosmic rays \cite{kostelecky:2015db}.

For the matter-gravity coefficients  ${\bar c}_{\mu \nu}$ and ${\bar a}_\mu$,  which break Lorentz symmetry and EEP, several experiments can be used for analysis in addition to the ones already mentioned above including ground-based gravimeter and WEP experiments. Dedicated satellite EEP tests are among the most sensitive where the relative acceleration  of two test bodies of different composition is the observable of interest. Upon relating the satellite frame coefficients to the standard Sun-centered frame used for the SME,  oscillations in the acceleration of the two masses  occur at a number of different harmonics of the satellite orbital and rotational frequencies,  as well as the Earth's orbital frequency.  Future tests of particular interest include the currently flying MicroSCOPE experiment \cite{touboul:2001kx,touboul:2012cr}.

While the focus of the discussion to follow are the results for the minimal SME coefficients $\bs^{\mu\nu}$, recent work has also involved the nonminimal SME coefficients in the pure-gravity sector associated with mass dimension $5$ and $6$ operators. One promising testing ground for these coefficients is sensitive short-range gravity experiments. The Newtonian force between two test masses becomes modified in the presence of local Lorentz violation  by an anisotropic quartic force that is controlled by a subset of coefficients from the Lagrangian organized  as the totally symmetric $({\bar k}_{\rm eff})_{jklm}$,  which has dimensions of length squared \cite{bailey:2015fk}. This contains $14$ measurable quantities and any one short-range experiment is sensitive to $8$ of them. Two key experiments,  from Indiana University and Huazhong University of Science and Technology,  have both reported analysis in the literature \cite{long:2015kx,shao:2015uq} . A recent work combines the two analyses to place new limits on all $14$, a priori independent, $({\bar k}_{\rm eff})_{jklm}$ coefficients \cite{shao:2016aa}. Other higher mass dimension coefficients play a role in gravitational wave propagation \cite{kostelecky:2016nx} and gravitational \v Cerenkov radiation \cite{kostelecky:2015db}.

To conclude this section, we ask: what can be said about the possible sizes of the coefficients for Lorentz violation? A broad class of hypothetical effects is described by the SME effective field theory framework,  but it is a test framework and as such does not make specific predictions concerning the sizes of these coefficients. One intriguing suggestion is that there is room in nature for violations of spacetime symmetry that are large compared to other sectors due to the intrinsic weakness of gravity. Considering the current status of the coefficients $\bs^{\mu\nu}$,  the best laboratory limits are at the $10^{-10}$-$10^{-11}$ level, with improvements of four orders of magnitude in astrophysical tests  on these coefficients \cite{kostelecky:2015db}.
However,  the limits are at the $10^{-8} \, {\rm m}^2$ level for the mass dimension $6$ coefficients  $({\bar k}_{\rm eff})_{jklm}$ mentioned above. Comparing this to the Planck length $10^{-35} \, {\rm m}$,  we see that symmetry breaking effects could still have escaped detection that are not Planck suppressed. This kind of ``countershading'' was first pointed out for the ${\bar a}_\mu$ coefficients \cite{kostelecky:2009jk}, which,  having dimensions of mass, can still be as large as a fraction of the electron mass and still lie within current limits. 

In addition, any action-based model that breaks local Lorentz symmetry either explicitly or spontaneously can be matched to a subset of the SME coefficients.
Therefore,  constraints on SME coefficients can directly constrain these models. Matches between various toy models and coefficients in the SME have been achieved for models that produce effective $\bs^{\mu\nu}$,  ${\bar c}_{\mu\nu}$, ${\bar a}_\mu$, and other coefficients. This includes vector and tensor field models of spontaneous Lorentz-symmetry breaking \cite{bailey:2006uq, seifert:2009rr, kostelecky:2005uq,kostelecky:2009kx, altschul:2010fk, kostelecky:2011kx},  models of quantum gravity \cite{gambini:1999aa, kostelecky:2009yu} and noncommutative quantum field theory \cite{carroll:2001aa}. Furthermore, Lorentz violations may also arise in the context of string field theory models \cite{kostelecky:2001fr}.

\section{Postfit analysis versus full modeling}\label{sec:postfit}
Since the last decade, several studies aimed to find upper limits on SME coefficients in the gravitational sector. A lot of these studies are based on the search of possible signals in post-fit residuals of experiments. This was done with LLR \cite{battat:2007uq}, GPB \cite{bailey:2013kq}, binary pulsars \cite{shao:2014qd,shao:2014rc} or Solar System planetary motions \cite{iorio:2012zr,hees:2015sf}. However, two  new works focused on a direct fit to data with LLR \cite{bourgoin:2016yu} and VLBI~\cite{le-poncin-lafitte:2016yq}, which are more satisfactory.

Indeed, in the case of a post-fit analysis, a simple modeling of extra terms containing SME coefficients are least square fitted in the residuals, attempting to constrain the SME coefficients of a testing function in residual noise obtained from a pure GR analysis, where of course Lorentz symmetry is assumed. It comes out correlations between SME coefficients and other global parameters previously fitted (masses, position and velocity\ldots) cannot be assessed in a proper way. In others words, searching hypothetical SME signals in residuals, i.e. in noise, can lead to an overestimated formal error on SME coefficients, as illustrated in the case of VLBI \cite{le-poncin-lafitte:2016yq}, and without any chance to learn something about correlations with other parameters, as for example demonstrated in the case of LLR \cite{bourgoin:2016yu}. Let us consider the VLBI example to illustrate this fact. The VLBI analysis is described in Section~\ref{sec:vlbi}. Including the SME contribution within the full VLBI modeling and estimating the SME coefficient $\bs^{TT}$ altogether with the other parameters fitted in standard VLBI data reduction leads to the estimate $\bs^{TT}=(-5\pm 8)\times 10^{-5}$. A postfit analysis performed by fitting the SME contribution within the VLBI residuals obtained after a pure GR analysis leads to $\bs^{TT}=(-0.6\pm 2.1)\times 10^{-8}$ \cite{le-poncin-lafitte:2016yq}. This example shows that a postfit analysis can lead to results with overoptimistic uncertainties and one needs to be extremely careful when using such results. 

\section{Data analysis}\label{sec:data}
In this section, we will review the different measurements that have already been used in order to constrain the SME coefficients. The different analyses are based on quite different types of observations. In order to compare all the corresponding results, we need to report them in a canonical inertial frame. The standard canonical frame used in the SME framework is a Sun-centered celestial equatorial frame \cite{kostelecky:2002uq}, which is approximately inertial over the time scales of most observations. This frame is asymptotically flat and comoving with the rest frame of the Solar System. The cartesian coordinates related to this frame are denoted by capital letters
\begin{equation}
	X^\Xi=(cT,X^J)=(cT,X,Y,Z)\, .
\end{equation}
The $Z$ axis is aligned with the rotation axis of the Earth, while the $X$ axis points along the direction from the Earth to the Sun at vernal equinox. The origin of the coordinate time $T$ is given by the time when the Earth crosses the Sun-centered $X$ axis at the vernal equinox. These conventions are depicted in Figure 2 from \cite{bailey:2006uq}. 

In the following subsections, we will present the different measurements used to constrain the SME coefficients. Each subsection contains a brief description of the principle of the experiment, how it can be used to search for Lorentz symmetry violations, what are the current best constraints obtained with such measurements and eventually how it can be improved in the future. 

\subsection{Atomic gravimetry}\label{sec:ai}
The most sensitive experiments on Earth searching for Lorentz Invariance Violation (LIV) in the minimal SME gravity sector are gravimeter tests.  As Earth rotates, the signal recorded in a gravimeter, i.e. the apparent local gravitational acceleration $g$ of a laboratory test body, would be modulated in the presence of LIV in gravity. This was first noted by Nordtvedt and Will in 1972 \cite{nordtvedt:1972vn} and used soon after with gravimeter data to constrain preferred-frame effects in the PPN formalism \cite{warburton:1976ys,nordtvedt:1976vn} at the level of $10^{-3}$. 

 This test used a superconducting gravimeter, based on a force comparison (the gravitational force is counter-balanced by an electromagnetic force maintaining the test mass at rest).  While superconducting gravimeters nowadays reach the best sensitivity on Earth, force comparison gravimeters intrinsically suffer from drifts of their calibration factor (with e.g. aging of the system). Development of other types of gravimeters has evaded this drawback: free fall gravimeters. Monitoring the motion of a freely falling test mass, they provide an absolute measurement of $g$. State-of-the art free fall gravimeters use light to monitor the mass free fall. Beyond classical gravimeters that drop a corner cube, the development of atom cooling and trapping techniques and atom interferometry has led to a new generation of free fall gravimeters, based on a quantum measurement: atomic gravimeters.

Atomic gravimeters use atoms in gaseous phase as a test mass. The atoms are initially trapped with magneto-optical fields in vacuum, and laser cooled (down to 100~nK) in order to control their initial velocity (down to a few mm/s). The resulting cold atom gas, containing typically a million atoms, is then launched or dropped for a free fall measurement. Manipulating the electronic and motional state of the atoms with two counterpropagating lasers, it is possible to measure, using atom interferometry, their free fall acceleration with respect to the local frame defined by the two lasers \cite{bordee:1989fk}. This sensitive direction is aligned to be along the local gravitational acceleration noted $\hat{z}$; the atom interferometer then measures the phase $\varphi=k a^{\hat{z}} T^2$, where $T$ is half the interrogation time, $k\simeq2(2\pi/\lambda)$ with $\lambda$ the laser wavelength, and $a^{\hat{z}}$ is the  free fall acceleration along the laser direction. The free fall time is typically on the order of 500~ms, corresponding to a free fall distance of about a meter. A new ``atom preparation - free fall - detection'' cycle is repeated every few seconds. Each measurement is affected by white noise, but averaging leads to a typical sensitivity on the order of or below $10^{-9}$~$g$ \cite{farah:2014uq,hauth:2013kx,hu:2013vn}.

Such an interferometer has been used by H. M{\"u}ller \textit{et al.} in \citep{muller:2008kx} and K.~Y. Chung \textit{et al.} in \citep{chung:2009uq} for testing Lorentz invariance in the gravitational sector with Caesium atoms, leading to the best terrestrial constraints on the $\bs^{\mu\nu}$ coefficients. The analysis uses three data sets of respectively 2.5 days for the first two and 10 days for the third, stretched over 4 years, which allows one to observe sidereal and annual LIV signatures. The gravitational SME model used for this analysis can be found in \cite{muller:2008kx,chung:2009uq,bailey:2006uq}; its derivation will be summarized hereunder. Since the atoms in free fall are sensitive to the local phase of the lasers, LIV in the interferometer observable could also come from the pure electromagnetic sector. This contribution has been included in the experimental analysis in \cite{chung:2009uq}. Focusing here on the gravitational part of SME, we ignore it in the following.

The gravitational LIV model adjusted in this test restricts to modifications of the Earth-atom two-body gravitational interaction. The Lagrangian describing the dynamics of a test particle at a point on the Earth's surface can be approximated by a post-Newtonian series as developed in \citep{bailey:2006uq}. At the Newtonian approximation, the two bodies Lagrangian is given by
\begin{equation}
 \mathcal{L}= \dfrac{1}{2}mV^2 + G_N \dfrac{Mm}{R}\left( 1   + \frac{1}{2}\bs^{JK}_t \hat{R}^J \hat R^K - \frac{3}{2} \bs^{TJ} \frac{V^J}{c} -   \bs^{TJ} \hat{R}^J \frac{V^K}{c}  \hat{R}^k  \right)\, ,
\end{equation}
where $\bm R$ and $\bm V$ are the position and velocities expressed in the standard SME Sun-centered frame and $\hat {\bm R }= \bm R/R$ with $R=\left|\bm R\right|$. In addition, we have introduced the $G_N$ is the observed Newton constant measured by considering the orbital motion of bodies and defined by (see also \cite{bailey:2006uq,hees:2015sf} or Section IV of \cite{bailey:2013kq})
\begin{equation}\label{eq:GN}
	G_N=G\left(1+\frac{5}{3}\bs^{TT}\right)\, ,
\end{equation}
and the 3-dimensional traceless tensor
\begin{equation}\label{eq:traceless}
	\bar{s}^{J\!K}_t=\bar{s}^{J\!K}\!-\!\frac{1}{3}\bar{s}^{T\!T}\delta^{J\!K}\, .
\end{equation}

From this Lagrangian one can derive the equations of motion of the free fall mass in a laboratory frame (see the procedure in Section V.C.1. from \cite{bailey:2006uq}). It leads to the modified local acceleration in the presence of LIV \citep{bailey:2006uq} given by
\begin{equation}
\label{eq-a-AG}
a^{\hat{z}} = g\left(1-  \frac{1}{6} i_4 \bar{s}^{TT} + \frac{1}{2}i_4 \bar {s}^{\hat{z}\hat{z}}\right) - \omega_\oplus^2 R_\oplus \sin^2 \chi - g i_4 \bar {s}^{T \hat{z}} \beta^{\hat{z}}_\oplus - 3 g i_1 \bar{s}^{TJ} \beta^J_\oplus \, ,
\end{equation}
where $g=G_NM_\oplus/R_\oplus^2$, $\omega_\oplus$ is the Earth's angular velocity, $\beta_\oplus = \frac{V_\oplus}{c} \sim 10^{-4}$ is the Earth's boost, $R_\oplus$ is the Earth radius, $M_\oplus$ is the Earth mass and $\chi$ the colatitude of the lab whose reference frame's $\hat{z}$ direction is the sensitive axis of the instrument as previously defined here. This model includes the shape of the Earth through its spherical moment of inertia $I_\oplus$ which appears in $i_\oplus = \frac{I_\oplus}{M_\oplus R^2_\oplus}$, $i_1 = 1 + \frac{1}{3} i_\oplus $ and $i_4 = 1 - 3i_\oplus $. In \citep{chung:2009uq}, Earth has been approximated as spherical and homogeneous leading to $i_\oplus =  \frac{1}{2}$, $i_1 = \frac{7}{6}$ and $i_4 =  -\frac{1}{2}$.

The sensing direction of the experiment precesses around the Earth rotation axis with sidereal period, and the lab velocity varies with sidereal period and annual period. At first order in $V_\oplus$ and $\omega_\oplus$ and as a function of the SME coefficients, the LIV signal takes the form of a harmonic series with sidereal and annual base frequencies (denoted resp. $\omega_\oplus$ and $\Omega$) together with first harmonics. The time dependence of the measured acceleration $a^{\hat{z}}$ from Eq.~(\ref{eq-a-AG}) arises from the terms involving the $\hat z$ indices. It can be decomposed in frequency according to \cite{bailey:2006uq}
\begin{equation}
 \dfrac{\delta a^{\hat{z}}}{a^{\hat{z}}}= \sum_l C_l \cos\left( \omega_l t + \phi_l\right) +  D_l \sin\left(  \omega_l t + \phi_l \right).
\label{eqmodelAG}
\end{equation}
The model contains seven frequencies $l \in \left\lbrace  \Omega,\omega_\oplus, 2 \omega_\oplus, \omega_\oplus \pm \Omega, 2 \omega_\oplus \pm \Omega  \right\rbrace$. The 14 amplitudes $C_l$ and $D_l$ are linear combinations of 7 $\bs^{\mu\nu}$ components: $\bs^{JK}$, $\bs^{TJ}$  and $\bs^{XX}-\bs^{YY}$ which can be found in Table 1 of \cite{chung:2009uq} or Table IV from \cite{bailey:2006uq}.

In order to look for tiny departures from the constant Earth-atom gravitational interaction, a tidal model for $a^{\hat{z}}$ variations due to celestial bodies is removed from the data before fitting to Eq. (\ref{eqmodelAG}). This tidal model consists of two parts. One part is based on a numerical calculation of the Newtonian tide-generating potential from the Moon and the Sun at Earth's surface based on ephemerides. It uses here the Tamura tidal catalog \citep{tamura:1987hc} which gives the frequency, amplitude and phase of 1200 harmonics of the tidal potential. These arguments are used by a software (ETGTAB) that calculates the time variation of the local acceleration in the lab and includes the elastic response of Earth's shape to the tides, called ``solid Earth tides'', also described analytically e.g. by the DDW model \cite{dehant:1999mz}.  A previous SME analysis of the atom gravimeter data using only this analytical tidal correction had been done, but it led to a degraded sensitivity of the SME test \cite{muller:2008kx}. Indeed, a non-negligible contribution to $a^{\hat{z}}$ is not covered by this non-empirical tidal model: oceanic tide effects such as ocean loading, for which good global analytical models do not exist. They consequently need to be adjusted from measurements. For the second analysis, reported here, additional local tidal corrections fitted on altimetric data have been removed \citep{egbert:1994jk} allowing to improve the statistical uncertainty of the SME test by one order of magnitude.

After tidal subtraction, signal components are extracted from the data using a numerical Fourier transform (NFT). Due to the finite data length, Fourier components overlap, but the linear combinations of spectral lines that the NFT estimates can be expressed analytically. Since the annual component $\omega_l=\Omega$ has not been included in this analysis, the fit provides 12 measurements. From there, individual constraints on the 7 SME coefficients and their associated correlation coefficients can be estimated by a least square adjustment. The results obtained are presented in Table \ref{tab:AI}.

\begin{table}[H]
\caption{Atom-interferometry limits on Lorentz violation in gravity from \cite{chung:2009uq}. The correlation coefficients can be derived from Table III of \cite{chung:2009uq}.}
\label{tab:AI}
\small 
\centering
\begin{tabular}{lr}
\toprule
\textbf{Coefficient}	& \\
\midrule
$ \bs^{TX}$			    & $\left(-3.1 \pm 5.1\right)  \times  10^{-5}$\\
$ \bs^{TY}$			    & $\left( 0.1 \pm 5.4\right)  \times  10^{-5}$\\
$ \bs^{TZ}$			    & $\left( 1.4 \pm 6.6\right)  \times  10^{-5}$\\
$ \bs^{XX}-\bs^{YY}$	& $\left( 4.4 \pm 11 \right)  \times  10^{-9}$\\
$ \bs^{XY}$          	& $\left( 0.2 \pm 3.9\right)  \times  10^{-9}$\\
$ \bs^{XZ}$          	& $\left(-2.6 \pm 4.4\right)  \times  10^{-9}$\\
$ \bs^{YZ}$          	& $\left(-0.3 \pm 4.5\right)  \times  10^{-9}$\\
\bottomrule
\end{tabular}\hspace{1cm}
\begin{tabular}{c c c c c c c}\toprule \multicolumn{7}{c}{\textbf{Correlation coefficients}}\\ \midrule
     1              &  \phantom{$10^{-4}$}  \\  
	\phantom{-}0.05 & 1                 &  \phantom{$10^{-4}$}\\ 
    \phantom{-}0.11 & -0.16             & 1                 & \phantom{$10^{-4}$} \\ 
  	         -0.82  & \phantom{-}0.34   & -0.16             &   1             & \phantom{$10^{-4}$} \\ 
             -0.38  & -0.86             & \phantom{-}0.10   & -0.01           & 1  & \phantom{$10^{-4}$}\\ 
           -0.41    & \phantom{-}0.13   & -0.89             & \phantom{-}0.38 & \phantom{-}0.02       & 1 &\phantom{$10^{-4}$}\\ 
           -0.12    & -0.19             & -0.89             & \phantom{-}0.04 & \phantom{-}0.20       & \phantom{-}0.80 & 1\phantom{$^{4}$}\\ 
\bottomrule
\end{tabular}
\end{table}

All results obtained are compatible with null Lorentz violation. As expected from boost suppressions in Eq. (\ref{eq-a-AG}) and from the measurement uncertainty, on the order of a few $10^{-9}~g$ \cite{peters:2001aa}, typical limits obtained are in the $10^{-9}$ range for purely spatial $\bs^{\mu\nu}$ components and 4 orders of magnitude weaker for the spatio-temporal components $\bs^{TJ}$. It can be seen e.g. with the purely spatial components that these constraints do not reach the intrinsic limit of acceleration resolution of the instrument (which has a short term stability of $11\times10^{-9}~g/\sqrt{\mbox{Hz}}$) because the coefficients are still correlated. Their marginalized uncertainty is broadened by their correlation.

Consequently, improving the uncertainty could be reached through a better decorrelation, by analyzing longer data series. In parallel, the resolution of these instruments keeps increasing and has nowadays improved by about a factor 10 since this experiment. However, increasing the instrument's resolution brings back to the question of possible accidental cancelling in treating ``postfit''  data. Indeed, it should be recalled here that local tidal corrections subtracted prior to analysis are based on adjusting a model of ocean surface from altimetry data. In principle, this observable would as well be affected by gravity LIV; fitting to these observations thus might remove part of SME signatures from the atom gravimeter data. This was mentioned in the first atom gravimeter SME analysis \cite{muller:2008kx}. The adjustment process used to assess local corrections in gravimeters is not made directly on the instrument itself, but it always involves a form of tidal measurement (here altimetry data, or gravimetry data from another instrument in \cite{merlet:2008fv}). All LIV frequencies match to the main tidal frequencies. Further progress on SME analysis with atom gravimeters would thus benefit from addressing in more details the question of possible signal canceling.

\subsection{Very Long Baseline Interferometry}\label{sec:vlbi}
VLBI is a geometric technique measuring the time difference in the arrival of a radio wavefront, emitted by a distant quasar, between at least two Earth-based radio-telescopes. VLBI observations are done daily since 1979 and the database contains nowadays almost 6000 24hours sessions, corresponding to 10 millions group-delay observations, with a present precision of a few picoseconds. One of the principal goals of VLBI observations is the kinematical monitoring of Earth rotation with respect to a global inertial frame realized by a set of defining quasars, the International Celestial Reference Frame \cite{fey:2015kq}, as defined by the International Astronomical Union \cite{soffel:2003bd}. The International VLBI Service for Geodesy and Astrometry (IVS)  organizes sessions of observation, storage of data and distribution of products, in particular the Earth Orientation parameters. Because of this precision, VLBI is also a very interesting tool to test gravitation in the Solar System. Indeed, the gravitational fields of the Sun and the planets are responsible of relativistic effects on  the quasar light beam through the propagation of the signal to the observing station and VLBI is able to detect these effects very accurately. By using the complete VLBI observations database, it was possible to obtain a constraint on the $\gamma$ PPN parameter  at the level of $1.2\times 10^{-4}$ \cite{lambert:2009bh,lambert:2011yu}. In its minimal gravitational sector, SME can also be  investigated with VLBI and obtaining a constrain on the $\bs^{TT}$ coefficient is possible. 

Indeed, the propagation time of a photon emitted at the event $(cT_e,\bm X_e)$ and received at the position $\bm X_r$ can be computed in the SME formalism using the time transfer function formalism \cite{le-poncin-lafitte:2004cr,teyssandier:2008nx,le-poncin-lafitte:2008fk,hees:2014fk,hees:2014nr} and is given by \cite{bailey:2006uq,bailey:2009fk}
\begin{align}      
&	\mathcal T (\bm X_e,T_e,\bm X_r)=T_r-T_e=\frac{R_{er}}{c}+2\frac{G_NM}{c^3}\left[1-\frac{2}{3}\bs^{TT}-\bs^{TJ}N_{er}^J\right] \ln \frac{R_e-\bm N_{er}.\bm X_e}{R_r-\bm N_{er}.\bm X_r}\label{eq:delay} \\
	&  +\frac{G_NM}{c^3} \Big(\bs^{TJ}P_{er}^J - \bs^{JK}N_{er}^J P_{er}^K\Big)\frac{R_e-R_r}{R_e R_r} +\frac{G_NM}{c^3} \Big\lbrack\bs^{TJ} N_{er}^J +\bs^{JK}\hat P_{er}^J\hat P_{er}^K-\bs^{TT}\Big\rbrack\left(\bm N_r . \bm N_{er}-\bm N_e . \bm N_{er}\right) \nonumber
 \end{align}
  where the terms $a_1$ and $a_2$ from \cite{bailey:2009fk} are taken as unity (which corresponds to using the harmonic gauge, which is the one used for VLBI data reduction), $R_e=\left|\bm X_e\right|$, $R_r=\left|\bm X_r\right|$, $R_{er}=\left|\bm X_r-\bm X_e\right|$ with the central body located at the origin and where we introduce the following vectors
\begin{equation}
	\bm K=\frac{\bm X_e}{R_e}\, , \quad	\bm N_{ij}\equiv\frac{\bm X_{ij}}{R_{ij}}=\frac{\bm X_j-\bm X_i}{|\bm X_{ij}|}\, , \quad
	\bm N_i=\frac{\bm X_i}{|\bm X_i|}\, , \quad \bm P_{er}=\bm N_{er}\times(\bm X_r\times \bm N_{er})  , \quad \textrm{and} \quad \hat {\bm P}_{er}=\frac{\bm P_{er}}{|\bm P_{er}|} \, ,
\end{equation}
and where $G_N$ is the observed Newton constant measured by considering the orbital motion of bodies and is defined in Eq.~(\ref{eq:GN}). This equation is the generalization of the well-known Shapiro time delay including Lorentz violation. The VLBI is actually measuring the difference of the time of arrival of a signal received by two different stations. This observable is therefore sensitive to a differential time delay (see \cite{finkelstein:1983gf} for a calculation in GR). Assuming a radio-signal emitted by a quasar at event $(T_e,\bm X_e)$ and received by two different VLBI stations at events $(T_1,\bm X_1)$ and $(T_2,\bm X_2)$ (all quantities being expressed in a barycentric reference frame), respectively, the VLBI group-delay $\Delta \tau_{(\textrm{SME})}$ in SME formalism can be written \cite{le-poncin-lafitte:2016yq} 
\begin{eqnarray}
\Delta \tau_{(\textrm{SME})}&=&2\frac{{G_NM}}{c^3}(1-\frac{2}{3}\bs^{TT}) \ln \frac{R_1+\bm K.\bm X_1}{R_2+\bm K.\bm X_2} +\frac{2}{3}\frac{{G_NM}}{c^3} \bs^{TT}\left(\bm N_2 . \bm K-\bm N_1 . \bm K \right)\, ,\label{eq:vlbi_stt}
\end{eqnarray}
where we only kept the $\bar s^{TT}$ contribution (see Eq.~(7) from \cite{le-poncin-lafitte:2016yq} for the full expression) and we use the same notations as in \cite{finkelstein:1983gf} by introducing three unit vectors
\begin{equation}
\bm K=\frac{\bm X_e}{|\bm X_e|}\, , \quad \bm N_1=\frac{\bm X_1}{|\bm X_1|}\, , \quad \textrm{and} \quad
\bm N_2=\frac{\bm X_2}{|\bm X_2|}\, .
\end{equation}

Ten million VLBI delay observations between August 1979 and mid-2015 have been used to estimate the $\bar s^{TT}$ coefficient.  First, VLBI observations are corrected from delay due to the radio wave crossing of dispersive media by using 2~GHz and 8~GHz recordings. Then, we used only the 8~GHz delays and the Calc/Solve geodetic VLBI analysis software, developed at NASA Goddard Space Flight Center and coherent with the latest standards of the International Earth Rotation and Reference Systems Service~\cite{petit:2010fk}. We added the partial derivative of the VLBI delay with respect to $\bar s^{TT}$ from Eq.~(\ref{eq:vlbi_stt}) to the software package using the USERPART module of Calc/Solve. We turned to a global solution in which we estimated $\bar s^{TT}$ as a global parameter together with radio source coordinates. We obtained 
\begin{equation}
	\bar s^{TT}=(-5\pm8)\times 10^{-5}\, , \label{eq:vlbi}
\end{equation}
with a postfit root mean square of 28~picoseconds and a $\chi^2$ per degree of freedom of 1.15. Correlations between radio source coordinates and $\bar s^{TT}$ are lower than 0.02, the global estimate being consistent with the mean value obtained with the session-wise solution with a slightly lower error.

In conclusion, VLBI is an incredible tool to test Lorentz symmetry, especially the $\bs^{TT}$ coefficient. This coefficient has an isotropic impact on the  propagation speed of gravitational waves as can be noticed from Eq.~(\ref{eq:speed}) below (or see Eq.~(9) from \cite{kostelecky:2015db} or Eq.~(11) from \cite{kostelecky:2016nx}). The analysis performed in \cite{le-poncin-lafitte:2016yq} includes the SME contribution in the modeling of VLBI observations and includes the $\bs^{TT}$ parameter in the global fit with other parameters. It is therefore a robust analysis that produces the current best estimate on the $\bar s^{TT}$ parameter. In the future, the accumulation of VLBI data in the framework of the permanent geodetic monitoring program leads us expect improvement of this constraint.

\subsection{Lunar Laser Ranging}\label{sec:llr}
On August, 20th 1969, after ranging to the lunar retro-reflector placed during the Apollo 11 mission, the first LLR echo was detected at the McDonald Observatory in Texas. Currently, there are five stations spread over the world which have realized laser shots on five lunar retro-reflectors. Among these stations four are still operating:  Mc Donald Observatory in Texas, Observatoire de la C\^ote d'Azur in France, Apache point Observatory in New Mexico and Matera in Italy while one on Maui, Hawaii has stopped lunar ranging since 1990. Concerning the lunar retro-reflectors three are located at sites of the Apollo missions 11, 14 and 15 and two are French-built array operating on the Soviet roving vehicle Lunakhod 1 and 2.

LLR is used to conduct high precision measurements of the light travel time of short laser pulses emitted at time $t_1$ by a LLR station, reflected at time $t_2$ by a lunar retro-reflector and finally received at time $t_3$ at a station receiver. The data are presented as normal points which combine time series of measured light travel time of photons, averaged over several minutes to achieve a higher signal-to-noise ratio measurement of the lunar range at some characteristic epoch. Each normal-point is characterized by one emission time ($t_1$ in universal time coordinate -- UTC), one time delay ($\Delta t_c$ in international atomic time -- TAI) and some additional observational parameters as laser wavelength, atmospheric temperature and pressure \emph{etc}. According to \cite{chapront:1999aa}, the theoretical pendent of the observed time delay ($\Delta t_c=t_3-t_1$ in TAI) is defined as 
\begin{equation}
  \Delta t_c=\Big[T_3-\Delta\tau_t(T_3)\Big]-\Big[T_1-\Delta\tau_t(T_1)\Big]\text{,}
  \label{eq:timdel}
\end{equation}
where $T_1$ is the emission time expressed in barycentric dynamical time (TDB)  and $\Delta\tau_t$ is a relativistic correction between the TDB and the terrestrial time (TT) at the level of the station. The reception time $T_3$ expressed in TDB is defined by the following two relations
\begin{subequations}\label{eq:timrecrep}
  \begin{align}
    T_3&=T_2+\frac{1}{c}\big\|\bm X_{o'}(T_3)-\bm X_{r}(T_2)\big\|+\Delta\mathcal T_{(\textrm{grav})}+\Delta\tau_a\text{,}\label{eq:timrec}\\
    T_2&=T_1+\frac{1}{c}\big\|\bm X_{r}(T_2)-\bm X_{o}(T_1)\big\|+\Delta\mathcal T_{(\textrm{grav})}+\Delta\tau_a\text{,}\label{eq:timref}
  \end{align}
\end{subequations}
with $T_2$ the time in TDB at the reflection point  $\bm X_{o}$ and $\bm X_{o'}$ are respectively the barycentric position vector at the emitter and the reception point, $\bm X_{r}$ is the barycentric position vector at the reflection point, $\Delta\mathcal T_{(\textrm{grav})}$ is the one way gravitational time delay correction and $\Delta\tau_a$ is the one way tropospheric correction.

LLR measurements are used to produce the Lunar ephemeris but also provide a unique opportunity to study the Moon's rotation, the Moon's tidal acceleration, the lunar rotational dissipation, etc \cite{dickey:1994zl}. In addition, LLR measurements have turn the Earth-Moon system into a laboratory to study fundamental physics and to conduct tests of the gravitation theory. Nordtvedt was the first to suggest that LLR can be used to test GR by testing one of its pillar: the Strong Equivalence Principle \cite{nordtvedt:1968uq,nordtvedt:1968ys,nordtvedt:1968zm}. He showed that precise laser ranging to the Moon would be capable of measuring precisely the ratio of gravitational mass to inertial mass of the Earth to an accuracy sufficient to constrain a hypothetical dependence of this ratio on the gravitational self-energy. He concluded that such a measurement could be used to test Einstein's theory of gravity and others alternative theories as scalar tensor theories. The best current test of the Strong Equivalence Principle is provided by a combination of torsion balance measurements with LLR analysis and is given by \cite{williams:2004ys,williams:2009ys,merkowitz:2010fk}
\begin{equation}
  \eta=(4.4\pm 4.5)\times 10^{-4}\, ,
  \label{eq:SEP}
\end{equation}
where $\eta$ is the Nordtvedt parameter that is defined as $m_G/m_I=1+\eta U/mc^2$ with $m_G$ the gravitational mass, $m_I$ the inertial mass and $U$ the gravitational self-energy of the body. Using the Cassini constraint on the $\gamma$ PPN parameter \cite{bertotti:2003uq} and the relation $\eta=4\beta-\gamma-3$ leads to a constraint on $\beta$ PPN parameter at the level $\beta-1=(1.2\pm1.1)\times10^{-4}$  \cite{williams:2009ys}.

In addition to tests of the Strong Equivalence Principle, many other tests of fundamental physics were performed with LLR analysis. For instance, LLR data can be used to search for a temporal evolution of the gravitational constant $\dot{G}/G$ \cite{williams:2004ys} and to constrain the fifth force parameters \cite{muller:2008aa}.  In addition, LLR has been used to constrain violation of the Lorentz symmetry in the PPN framework. \citet{muller:2008aa} deduced from LLR data analysis constraints on the preferred frame parameters $\alpha_1$ and $\alpha_2$ at the level $\alpha_1=(-7\pm 9)\times 10^{-5}$ and $\alpha_2=(1.8\pm 2.5)\times 10^{-5}$.

Considering all the successful GR tests performed with LLR observations, it is quite natural to use them to search for Lorentz violations in the gravitation sector. In the SME framework, \citet{battat:2007uq} used the lunar orbit to provide estimates on the SME coefficients. Using a perturbative approach, the main signatures produced by SME on the lunar orbit have analytically been computed in \cite{bailey:2006uq}. These computations give a first idea of the amplitude of the signatures produced by a breaking of Lorentz symmetry. Nevertheless, these analytical signatures have been computed assuming the lunar orbit to be circular and fixed (i.e. neglecting the precession of the nodes for example). These analytical signatures have been fitted to LLR residuals obtained from a data reduction performed in pure GR \cite{battat:2007uq}. They determined a ``\emph{realistic}'' error on their estimates from a similar postfit analysis performed in the PPN framework.  The results obtained by this analysis are presented in Table \ref{tab:battat}. It is important to note that this analysis uses projections of the SME coefficients into the lunar orbital plane $\bar s^{11}, \bar s^{22}, \bar s^{0i}$ (see Section V.B.2 of \cite{bailey:2006uq}) while the standard SME analyses uses coefficients defined in a Sun-centered equatorial frame (and denoted by capital letter $\bs^{IJ}$).

\begin{table}[H]
\caption{Estimation of SME coefficients from LLR postfit data analysis from \cite{battat:2007uq}. No correlations coefficients have been derived in this analysis. The coefficients $\bar s^{ij}$ are projections of the $\bar s^{IJ}$ into the lunar orbital plane (see Eq. (107) from \cite{bailey:2006uq}) while the linear combinations $\bar{s}_{\Omega_{\oplus}c}$ and $\bar{s}_{\Omega_{\oplus}s}$ are given by Eq. (108) from \cite{bailey:2006uq}.}
\label{tab:battat}
\small 
\centering
\begin{tabular}{lr}
\toprule
\textbf{Coefficient}	& \\
\midrule
$\bar{s}^{11}-\bar{s}^{22}$   & $(1.3\pm 0.9)\times 10^{-10}$\\
$\bar{s}^{12}$                & $(6.9\pm 4.5)\times 10^{-11}$\\
$\bar{s}^{01}$                & $(-0.8\pm 1.1)\times 10^{-6\ }$\\
$\bar{s}^{02}$                & $(-5.2\pm 4.8)\times 10^{-7\ }$\\
$\bar{s}_{\Omega_{\oplus}c}$  & $(0.2\pm 3.9)\times 10^{-7\ }$\\
$\bar{s}_{\Omega_{\oplus}s}$  & $(-1.3\pm 4.1)\times 10^{-7\ }$\\
\bottomrule
\end{tabular}
\end{table}

However, as discussed in Section~\ref{sec:postfit} and in \cite{le-poncin-lafitte:2016yq,bourgoin:2016yu}, a postfit search for SME signatures into residuals of a data reduction previously performed in pure GR is not fully satisfactory. First of all, the uncertainties obtained by a postfit analysis based on a GR data reduction can be underestimated by up to two orders of magnitude. This is mainly due to correlations between SME coefficients and others global parameters (masses, positions and velocities, $\ldots$) that are neglected in this kind of approach. 
In addition, in the case of LLR data analysis, the oscillating signatures derived in \cite{bailey:2006uq} and used in \cite{battat:2007uq} to determine pseudo-constraints are computed only accounting for short periodic oscillations, typically at the order of magnitude of the mean motion of the Moon around the Earth. Therefore, this analytic solution remains only valid for few years while LLR data spans over 45 years (see also the discussions in footnote 2 from \cite{hees:2015sf} and page 22 from \cite{bailey:2006uq}).

\par Regarding  LLR data analysis, a more robust strategy consists in including the SME modeling in the complete data analysis and to estimate the SME coefficients in a global fit along with others parameters by taking into account short and long period terms and also correlations (see \cite{bourgoin:2016yu}). In order to perform such an analysis, a new numerical lunar ephemeris named ``\'Eph\'em\'eride Lunaire Parisienne Num\'erique'' (ELPN) has been developed within the SME framework. The dynamical model of ELPN is similar to the DE430 one \cite{folkner:2014uq} but includes the Lorentz symmetry breaking effects arising on the orbital motion of the Moon. The SME contribution to the lunar equation of motion has been derived in \cite{bailey:2006uq} and is given by
\begin{align}
  a^J_{\text{SME}}=\ &\frac{G_N M}{r^3}\Big[\bar{s}^{J\!K}_tr^K\!-\frac{3}{2}\bar{s}^{K\!L}_t\hat{r}^K\hat{r}^Lr^J+2\frac{\delta m}{M}\Big(\bar{s}^{T\!K}\hat{v}^Kr^J-\bar{s}^{T\!J}\hat{v}^Kr^K\Big)\Big.\nonumber\\
  &\quad\Big.+3\bar{s}^{T\!K}\hat{V}^Kr^J-\bar{s}^{T\!J}\hat{V}^Kr^K-\bar{s}^{T\!K}\hat{V}^Jr^K+3\bar{s}^{T\!L}\hat{V}^K\hat{r}^K\hat{r}^Lr^J\Big]\text{,}\label{eq:eom}
\end{align}
where $G_N$ is the observed Newtonian constant defined by Eq.~(\ref{eq:GN}), $M$ is the mass of the Earth-Moon barycenter, $\delta m$ is the difference between the Earth and the lunar masses; $\hat{r}^J$ being the unit position vector of the Moon with respect to the Earth; $\hat{v}^J\!=\!v^J\!/c$ with $v^J$ being the relative velocity vector of the Moon with respect to the Earth; $\hat{V}^J\!=\!V^J\!/c$ with $V^J$ being the Heliocentric velocity vector of the Earth-Moon barycenter and the 3-dimensional traceless tensor defined by Eq.~(\ref{eq:traceless}). These equations of motion as well as their partial derivatives are integrated numerically in ELPN.

In addition to the orbital motion, effects of a violation of Lorentz symmetry on the light travel time of photons is also considered. More precisely, the gravitational time delay $\Delta \mathcal T_{(\textrm{grav})}$ appearing in Eq. (\ref{eq:timdel}) is given by the gravitational part of Eq.~(\ref{eq:delay}) \cite{bailey:2009fk}.

Estimates on the SME coefficients are obtained by a standard chi-squared minimization: the LLR residuals are minimized by an iterative weighted least squares fit using partial derivatives previously computed from variational equations in ELPN. After an adjustment of 82 parameters including the SME coefficients a careful analysis of the covariance matrix shows that LLR data does not allow to estimate independently all the SME coefficients but that they are sensitive to the following three linear combinations:
\begin{equation}
  \begin{array}{lclcl}
    \bs^{XX}-\bs^{YY}\, , & \ \ \ \ & \bs^{TY}+0.43\bs^{TZ}\text{,} & \ \ \ \ & \bs^{XX}+\bs^{YY}-2\bs^{ZZ}-4.5\bs^{YZ}\text{.}
  \end{array}
  \label{eq:lin}
\end{equation}
The estimations on the 6 SME coefficients derived in \cite{bourgoin:2016yu} is summarized in Table \ref{tab:bourgoin}. In particular, it is worth emphasizing that the quoted uncertainties are the sum of the statistical uncertainties obtained from the least-square fit with estimations of systematics uncertainties obtained with a Jackknife resampling method \cite{lupton:1993aa,gottlieb:2003aa}.
\begin{table}[H]
\caption{Estimation of SME coefficients from a full LLR data analysis from \cite{bourgoin:2016yu} and associated correlation coefficients.}
\label{tab:bourgoin}
\small 
\centering
\begin{tabular}{lr}
\toprule
\textbf{Coefficient}	& \textbf{Estimates}\\
\midrule
$\bar s^{TX}$                                 & $\left(-0.9\pm 1.0\right)\times 10^{-8\phantom{0}}$\\
$\bs^{XY}$			                          & $\left(-5.7 \pm 7.7\right)\times 10^{-12}$         \\
$ \bs^{XZ}$			                          & $\left(-2.2 \pm 5.9\right)\times 10^{-12}$         \\
$ \bs^{XX}-\bs^{YY}$                          & $\left(0.6 \pm 4.2\right)\times 10^{-11}$\\
$ \bs^{TY}+0.43~\bs^{TZ}$                     & $\left(6.2 \pm 7.9\right)\times 10^{-9\phantom{0}}$\\
$ \bs^{XX}+\bs^{YY}-2\bs^{ZZ}-4.5~\bs^{YZ}$ & $\left(2.3 \pm 4.5\right)\times 10^{-11}$\\
\bottomrule
\end{tabular}\hfill
\begin{tabular}{c c c c c c }\toprule \multicolumn{6}{c}{\textbf{Correlation coefficients}}\\ \midrule
1 &  \phantom{$10^{-1}$}  \\  
	-0.06          & 1  & \phantom{$10^{-1}$}\\ 
    -0.04          & \phantom{-}0.29   & 1    & \phantom{$10^{-1}$} \\ 
 \phantom{-}0.58   & -0.12             & -0.16   &   1 & \phantom{$10^{-1}$} \\ 
 \phantom{-}0.16   & -0.01             & -0.09   & \phantom{-}0.25 & 1  & \phantom{$10^{-1}$}\\ 
 \phantom{-}0.07   & -0.10             & -0.13   & -0.10           & 0.03       & 1\phantom{$^{1}$}\\ 
\bottomrule
\end{tabular}
\end{table}

In summary, LLR is a powerful experiment to constrain gravitation theory and in particular hypothetical violation of the Lorentz symmetry. A first analysis based on a postfit estimations of the SME coefficients have been performed \cite{battat:2007uq} which is not satisfactory regarding the neglected correlations with other global parameters as explained in Section~\ref{sec:postfit}.  A full analysis including the integration of the SME equations of motion and the SME contribution to the gravitational time delay has been done in~\cite{bourgoin:2016yu}. The resulting estimates on some SME coefficients are presented in Table~\ref{tab:bourgoin}. In addition, some SME coefficients are still correlated with parameters appearing in the rotational motion of the Moon as the principal moment of inertia, the quadrupole moment, the potential Stockes coefficient $C_{22}$ and the polar component of the velocity vector of the fluid core~\cite{bourgoin:2016yu}. A very interesting improvement regarding this analysis would be to produce a joint GRAIL (Gravity Recovery And Interior Laboratory) \cite{konopliv:2014aa,lemoine:2014aa,arnold:2015aa} and LLR data analysis that would help in decorrelating the SME parameters from the lunar potential Stockes coefficients of degree 2 and therefore improve marginalized estimations of the SME coefficients. Finally, in \cite{battat:2007uq} and \cite{bourgoin:2016yu}, the effects of SME on the translational lunar equations of motion are considered and used to derive constraints on the SME coefficients. It would be also interesting to extend these analyses by considering the modifications due to SME on the rotation of the Moon. A first attempt has been proposed in Section V. A. 2. of \cite{bailey:2006uq} but needs to be extended.

\subsection{Planetary ephemerides}\label{sec:planets}
The analysis of the motion of the planet Mercury around the Sun was historically the first evidence in favor of GR with the explanation of the famous advance of the perihelion in 1915. From there, planetary ephemerides have always been a very powerful tool to constrain GR and alternative theories of gravitation. Currently, three groups in the world are producing planetary ephemerides: the NASA Jet Propulsion Laboratory with the DE ephemerides~\cite{standish:1982uq,newhall:1983uq,standish:1990vn,standish:2010ly,standish:2012fk,folkner:2014uq,hees:2014jk}, the French INPOP (Int\'egrateur Num\'erique Plan\'etaire de l'Observatoire de Paris)  ephemerides~\cite{fienga:2008fk,fienga:2009kx,fienga:2010vn,fienga:2011qf,verma:2014jk,fienga:2015rm} and the Russian EPM ephemerides~\cite{pitjeva:2005kx,pitjeva:2010ys,pitjeva:2013fk,pitjeva:2013uq,pitjeva:2014fj}. These analyses use an impressive number of different observations to produce high accurate planetary and asteroid trajectories. The observations used to produce ephemerides comprise radioscience observations of spacecraft that orbited around Mercury, Venus, Mars and Saturn, flyby tracking of spacecraft close to Mercury, Jupiter, Uranus and Neptune and optical observations of all planets. This huge set of observations have been used to constrain the $\gamma$ and $\beta$ post-Newtonian parameter at the level of $10^{-5}$ \cite{konopliv:2011dq,verma:2014jk,fienga:2015rm,pitjeva:2013uq,pitjeva:2014fj}, the fifth force interaction (see ~\cite{talmadge:1988uq} and Figure~31 from \cite{konopliv:2011dq}), the quantity of Dark Matter in our Solar System~\cite{pitjev:2013qv}, the Modified Newtonian Dynamics \cite{milgrom:2009vn,blanchet:2011ys,hees:2014jk,hees:2016mi}, \dots

A violation of Lorentz symmetry within the gravity sector of SME  induces different types of effects that can have implications on planetary ephemerides analysis: effects on the orbital dynamics and effects on the light propagation. Simulations using the Time Transfer Formalism \cite{teyssandier:2008nx,hees:2014fk,hees:2014nr} based on the software presented in \cite{hees:2012fk} have shown that only the $\bs^{TT}$ coefficients produce a non-negligible effect on the light propagation (while it has impact only at the next post-Newtonian level on the orbital dynamics~\cite{bailey:2006uq,kostelecky:2011kx}). On the other hand, the other coefficients produce non-negligible effects on the orbital dynamics~\cite{bailey:2006uq} and can therefore be constrained using planetary ephemerides data. In the linearized gravity limit, the contribution from SME to the 2-body equations of motion within the gravitational sector of SME are given by the first line of Eq. (\ref{eq:eom}) (i.e. for a vanishing $V^k$). The coefficient $\bs^{TT}$ is completely unobservable in this context since absorbed in a rescaling of the gravitational constant (see the discussion in~\cite{bailey:2006uq,bailey:2013kq}).

Ideally, in order to perform a solid estimation of the SME coefficients using planetary ephemerides, one should include the full SME equations in the integration of the planets motion and fit them simultaneously with the other estimated parameters (positions and velocities of planets, $J_2$ of the Sun, \dots). This solid analysis within the SME formalism has not been performed so far. 

As a first step, a postfit analysis has been performed \cite{iorio:2012zr,hees:2015sf}. The idea of this analysis is to derive the analytical expression for the secular evolution of the orbital elements produced by the SME contribution to the equations of motion. Using the Gauss equations, secular perturbations induced by SME on the orbital elements have been computed  in  \cite{bailey:2006uq} (see also \cite{iorio:2012zr} for a similar calculations done for the $\bs^{TJ}$ coefficients only). In particular, the secular evolution of the longitude of the ascending node $\Omega$ and the argument of the perihelion $\omega$ is given by
\begin{subequations}\label{eq:SMEad}
	\begin{eqnarray}
		\left<\frac{d\Omega}{dt}\right>&=&\frac{n}{\sin i(1-e^2)^{1/2}}\left[\frac{\varepsilon}{e^2}\bs_{kP}\sin\omega    \ +\frac{(e^2-\varepsilon)}{e^2}\bs_{kQ}\cos\omega-\frac{\delta m}{M} \frac{2na\varepsilon}{ec}\bs^k \cos\omega\right]\, ,\\
	\left<\frac{d\omega}{dt}\right>&=&	-\cos i \left<\frac{d\Omega}{dt}\right> -n\left[-\frac{\varepsilon^2}{2e^4}(\bs_{PP}-\bs_{QQ})+\frac{\delta m}{M}\frac{2na(e^2-\varepsilon)}{ce^3(1-e^2)^{1/2}}\bs^Q	\right] \, ,
	\end{eqnarray}
\end{subequations}
where $a$ is the semimajor axis, $e$ the eccentricity, $i$ the orbit inclination (with respect to the ecliptic), $n=(G_Nm_\odot/a^3)^{1/2}$ is the mean motion, $\varepsilon=1-(1-e^2)^{1/2}$, $\delta m$ the difference between the two masses and $M$  their sum (in the cases of planets orbiting the Sun, one has $M\approx \delta m$). In all these expressions, the coefficients for Lorentz violation with subscripts $P$, $Q$, and $k$ are understood to be appropriate projections of $\bs^{\mu\nu}$ along the unit vectors $P$, $Q$, and $k$, respectively. For example, $\bs^k=k^i \bs^{Ti}$, $\bs_{PP}=P^iP^j \bs^{ij}$. The unit vectors $P$, $Q$ and $k$ define the orbital plane (see \cite{bailey:2006uq} or Eq.~(8) from \cite{hees:2015sf}).

Instead of including the SME equations of motion in planetary ephemerides, the postfit analysis uses estimations of supplementary advances of perihelia and nodes derived from ephemerides analysis \cite{pitjeva:2013fk,pitjev:2013qv,fienga:2011qf} to fit the SME coefficients through Eq.~(\ref{eq:SMEad}). In \cite{hees:2015sf},  estimations of supplementary advances of perihelia and longitude of nodes from INPOP (see Table 5 from \cite{fienga:2011qf}) are used to fit a posteriori the SME coefficients. This analysis suffers from large correlations due to the fact that the planetary orbits are very similar to each other: nearly eccentric orbit and very low inclination orbital planes. In order to deal properly with these correlations a Bayesian Monte Carlo inference has been used~\cite{hees:2015sf}. The posterior probability distribution function can be found on Figure~1 from \cite{hees:2015sf}. The intervals corresponding to the 68\% Bayesian confidence levels are given in Table~\ref{tab:planets} as well as the correlation matrix. It is interesting to mention that a decomposition of the normal matrix in eigenvectors  allows one to find linear combinations of SME coefficients that are uncorrelated with the planetary ephemerides analysis (see Eq.~(15) and Table IV from \cite{hees:2015sf}).

\begin{table}[htb]
\caption{Estimations of the SME coefficients from a postfit data analysis based on planetary ephemerides from \cite{hees:2015sf}. The uncertainties correspond to the 68\% Bayesian confidence levels of the marginal posterior probability distribution function. The associated correlation coefficients can be found in Table III from \cite{hees:2015sf}.}
\label{tab:planets}
\small 
\centering
\begin{tabular}{lr}
\toprule
\textbf{Coefficient}	& \\
\midrule
 $\bs^{XX}-\bs^{YY}$			 & $(-0.8 \pm 2.0)\times 10^{-10}$ \\
 $\bs^{XX}+\bs^{YY}-2~\bs^{ZZ} $ 	 & $(-0.8 \pm 2.7)\times 10^{-10}$\\
 $\bs^{XY}$        	 & $(-0.3 \pm 1.1)\times 10^{-10}$\\
 $\bs^{XZ}$ 		 & $(-1.0 \pm 3.5)\times 10^{-11}$\\
 $\bs^{YZ}$ 		 & $(5.5 \pm 5.2)\times 10^{-12}$\\
 $\bs^{TX}$	         & $(-2.9 \pm 8.3)\times 10^{-9\phantom{1}}$ \\
 $\bs^{TY}$		     & $(0.3 \pm 1.4)\times 10^{-8\phantom{1}}$ \\
 $\bs^{TZ}$ 		 & $(-0.2 \pm 5.0)\times 10^{-8\phantom{1}}$ \\
\bottomrule
\end{tabular}\hfill
\begin{tabular}{c c c c c c c c}\toprule \multicolumn{8}{c}{\textbf{Correlation coefficients}}\\ \midrule
1\\
 \phantom{-}0.99 & 1\\
 \phantom{-}0.99 & \phantom{-}0.99   & 1 &\phantom{$10^{-1}$} \\
 \phantom{-}0.98   & \phantom{-}0.98 & \phantom{-}0.99 &1&\phantom{$10^{-1}$}\\
	-0.32          & -0.24           & -0.26           & -0.26            &1&\phantom{$10^{-1}$}\\
 \phantom{-}0.99  & \phantom{-}0.98  & \phantom{-}0.98 & \phantom{-}0.98  & -0.32            &1&\phantom{$10^{-1}$}\\
 \phantom{-}0.62  & \phantom{-}0.67  & \phantom{-}0.62  & \phantom{-}0.59 & \phantom{-}0.36  & \phantom{-}0.60&1&\phantom{$10^{-1}$}\\
  -0.83          & -0.86             & -0.83           & -0.81            & -0.14             &-0.82 & -0.95 &1\phantom{$^{-1}$}\\
\bottomrule
\end{tabular}
\end{table}

In summary, planetary ephemerides offer a great opportunity to constrain hypothetical violations of Lorentz symmetry. So far, only postfit estimations of the SME coefficients have been performed~\cite{iorio:2012zr,hees:2015sf}. In this analysis, estimations of secular advances of perihelia and longitude of nodes obtained with the INPOP planetary ephemerides \cite{fienga:2011qf} are used to fit a posteriori the SME coefficients using the Eqs.~(\ref{eq:SMEad}). The 68\% marginalized confidence intervals are given in Table~\ref{tab:planets}. This analysis suffers highly from correlations due to the fact that the planetary orbits are very similar. A very interesting improvement regarding this analysis would be to perform a full analysis by integrating the planetary equations of motion directly within the SME framework and by fitting the SME coefficients simultaneously with the other parameters fitted during the ephemerides data reduction.

\subsection{Gravity Probe B}\label{sec:gpb}
In GR, a gyroscope in orbit around a central body undergoes two relativistic precessions with respect to a distant inertial frame: (i) a geodetic drift in the orbital plane due to the motion of the gyroscope in the curved spacetime~\cite{de-sitter:1916cr} and (ii) a frame-dragging due to the spin of the central body~\cite{lense:1918fk}. In GR, the spin of a gyroscope is parallel transported, which at the post-Newtonian approximation gives the relativistic drift
\begin{subequations}
\begin{align}
    \bm  R &= \frac{d \hat {\bm  S}}{dt}=\bm  \Omega_{GR}\times \bm S\, ,  \label{eq:precession} \\
     \bm \Omega_{GR}&=\frac{3GM}{2c^2r^3}\bm r \times \bm v + \frac{3 \hat {\bm r} (\hat{\bm r}.\bm J)-\bm J}{c^2 r^3}\, ,\label{eq:prec_GR}
\end{align}
\end{subequations}
where $\hat{\bm S}$ is the unit vector pointing in the direction of the spin $\bm S$ of the gyroscope, $\bm r$ and $\bm v$ are the position and velocity of the gyroscope, $\hat {\bm r}=\bm r/r$ and $\bm J$ is the angular momentum of the central body.  In 1960, it has been suggested to use these two effects to perform a new test of GR~\cite{schiff:1960fk,pugh:1959rw}. In April 2004, GPB, a satellite carrying 4 cryogenic gyroscopes was launched in order to measure these two precessions. GPB was orbiting Earth on a polar orbit such that the two relativistic drifts are orthogonal to each other~\cite{everitt:2011fk}: the geodetic effect is directed along the NS direction (North-South, i.e. parallel to the satellite motion) while the frame-dragging effect is directed on the WE direction (West-East, see \cite{everitt:2011fk,bailey:2013kq} for further details about the axes conventions in the GPB data reduction). A year of data gives the following measurements of the relativistic drift: (i) the geodetic drift $R_{NS}=-6601.8\pm 18.3$ mas/yr (milliarcsecond per year) to be compared to the GR prediction of -6606.1 mas/yr and (ii) the frame-dragging drift $R_{WE}=-37.2\pm 7.2$ mas/yr to be compared with the GR prediction of -39.2 mas/yr. In other word, the GPB results can be written as a measurement of a deviation from GR given by 
\begin{equation}\label{eq:GPB}
    \Delta R_{NS}=4.3 \pm 18.3 \textrm{ mas/yr}\qquad \textrm{and}\qquad \Delta R_{WE}=2\pm 7.2 \textrm{ mas/yr}\, .
\end{equation}

Within the SME framework, if one considers only the $\bs^{\mu\nu}$ coefficients, the equation of parallel transport in term of the spacetime metric is not modified (see Eq.~(143) from \cite{bailey:2006uq}). Nevertheless, the expression of the spacetime metric is modified leading to a modification of the relativistic drift given by Eq. (150) from \cite{bailey:2006uq}. In order to focus only on the dominant secular part of the evolution of the spin orientation, the relativistic drift equation has been averaged over a period. The SME contribution to the precession can be written as \cite{bailey:2006uq}
\begin{equation}
    \Delta \Omega^J = \frac{G_NM}{r^2}v\left[\left(-\frac{4}{3}\bs^{TT}-\frac{9}{8}\tilde i_{(-5/3)}\bs^{JK}_t \hat\sigma^J\hat\sigma^K\right)\hat\sigma^J  +\frac{5}{4}\tilde i_{(-3/5)} \bs^{JK}_t \hat\sigma^K\right]\, ,
\end{equation}
where $G_N$ is the effective gravitational constant defined by Eq.~(\ref{eq:GN}), the coefficients $\tilde i$ are defined by $\tilde i_{(\beta)}=1+\beta I_\oplus/(M_\oplus r^2)$, $\hat \sigma^J$ is a unit vector normal to the gyroscope orbital plane, $r$ and $v$ are the norm of the position and velocity of the gyroscope and $\bs^{JK}_t$ is the traceless part of $\bar s^{JK}$ as defined by Eq.~(\ref{eq:traceless}). Using the geometry of GPB into the last equation and using Eq.~(\ref{eq:precession}), one finds that the gyroscope anomalous drift is given by
\begin{subequations}\label{eq:drift1}
\begin{align}
    \Delta R_{NS}&=5872\bs^{TT} + 794 \left(\bs^{XX}-\bs^{YY}\right)-317 \left(\bs^{XX}+\bs^{YY}-2\bs^{ZZ}\right)-1050\bs^{XY}\, ,\\
    \Delta R_{WE}&=-368 (\bs^{XX}-\bs^{YY}) -1112\bs^{XY}+1269\bs^{XZ}+4219\bs^{YZ}\, ,
\end{align}
\end{subequations}
where the units are mas/yr.
These are the SME modifications to the relativistic drift arising from the modification of the equations of evolution of the gyroscope axis (i.e. modification of the parallel transport equation due to the modification of the underlying spacetime metric). 

In addition to modifying the evolution of the spin axis, a breaking of Lorentz symmetry will impact the orbital motion of the gyroscope. As a result, the position and velocity of the gyroscope will depend on the SME coefficients and therefore also impact the evolution of the spin axis through the GR contribution given by Eq.~(\ref{eq:prec_GR}). The best way to deal with this effect is to use the GPB tracking measurements (GPS) in order to constrain the gyroscope orbital motion and eventually constrain the SME coefficients through the equations of motion. In \cite{bailey:2013kq}, these tracking observations are not used and only the gyroscope drift is used in order to constrain the SME contributions coming from both the modification of the parallel transport and from the modification of GPB orbital motion. In order to do this, the contribution of SME on the evolution of the orbital elements  given by Eq.~(\ref{eq:SMEad}) and~(\ref{eq:SMEad2}) are used, averaged over a period and in the low eccentricity approximation. This secular evolution for the osculating elements is introduced in the relativistic drift equation for the gyroscope from Eq.~(\ref{eq:prec_GR}) and averaged over the measurement time using Eq.~(\ref{eq:precession}). Using the GPB geometry, this contribution to the relativistic drift is given by
\begin{subequations}\label{eq:drift2}
\begin{align}
    \Delta R'_{NS}&=  5.7\times 10^{6} (\bs^{XX}-\bs^{YY}) + 1.7\times 10^{7}\bs^{XY} - 1.9\times 10^7 \bs^{XZ} -6.6 \times 10^7 \bs^{YZ}\, ,\\
    \Delta R'_{WE}&=-1.89\times 10^{7} (\bs^{XX}-\bs^{YY}) -5.71 \times 10^7 \bs^{XY}- 5.96 \times 10^6 \bs^{XZ}  -1.98\times 10^7 \bs^{YZ} \, ,
\end{align}
\end{subequations}
with units of mas/yr.

The sum of the two SME contributions to the gyroscope relativistic drift given by Eq.~(\ref{eq:drift1}) and (\ref{eq:drift2}) can be compared to the GPB estimations given by Eq.~(\ref{eq:GPB}). The result is given in Table~\ref{tab:GPB}. The main advantage of GPB comes from the fact that it is sensitive to the $\bs^{TT}$ coefficient. The constraint on this coefficient is at the level of $10^{-3}$, a little bit less good than the one obtained with VLBI or with binary pulsars but relying on a totally different type of observations. The constraints on the spatial part of the SME coefficients ($\bs^{IJ}$) are at the level of $10^{-7}$ and are superseded by the other measurements. The constraints on these coefficients come mainly from the contribution arising from the orbital dynamics of GPB and not from a direct modification of the spin evolution. Constraining the orbital motion from GPB by using the gyroscope observations only is not optimal and tracking observations may help to improve the corresponding constraints (in this case, a dedicated satellite may be more appropriate as discussed in Section \ref{sec:slr}). 
\begin{table}[htb]
\caption{Estimations of the SME coefficients from a postfit data analysis based on GPB \cite{bailey:2013kq}.}
\label{tab:GPB}
\small 
\centering
\begin{tabular}{lr}
\toprule
\textbf{Coefficient}	& \\
\midrule
 $\bs^{(1)}_\textrm{GPB}=\bs^{TT}+970\left(\bs^{XX}-\bs^{YY}\right) -0.05 \left(\bs^{XX}+\bs^{YY}-2\bs^{ZZ}\right)$ &  \\
 $\hspace{5cm}+2895\, \bs^{XY} - 3235\, \bs^{XZ}-11\, 240 \,\bs^{YZ}$ & $(0.7 \pm 3.1)\times 10^{-3}$\\
 $\bs^{(2)}_\textrm{GPB}=\bs^{XX}-\bs^{YY}+3.02\, \bs^{XY}+0.32\, \bs^{XZ}+1.05 \, \bs^{YZ}$ 	 & $(-1.1 \pm 3.8)\times 10^{-7}$\\
 \bottomrule
\end{tabular}
\end{table}

In summary, the GPB measurement of a gyroscope relativistic drifts due to geodetic precession or frame-dragging can be used to search for a breaking of Lorentz symmetry. The main advantage of this technique comes from its sensitivity to $\bs^{TT}$. As already mentioned, this coefficient has an isotropic impact on the propagation velocity of gravitational waves as can be noticed from Eq.~(\ref{eq:speed}) below (see also Eq.~(9) from \cite{kostelecky:2015db} or Eq.~(11) from \cite{kostelecky:2016nx}). A preliminary result based on a post-fit analysis performed after a GR data reduction of GPB measurements gives a constraint on $\bs^{TT}$ at the level of $10^{-3}$ \cite{bailey:2013kq}. This should be investigated further since the Earth's quadrupole moment has been neglected and Lorentz-violating effects on the aberration terms can also change slightly the results. In addition, impacts from Lorentz violations on frame-dragging arising in other contexts such as satellite laser ranging (see Section~\ref{sec:slr}) or signals from accretion disks around collapsed stars \cite{stella:1999ek} would also be interesting to consider.

\subsection{Binary pulsars}\label{sec:pulsars}
The discovery of the first binary pulsars PSR 1913+16 by Hulse and Taylor in 1975 \cite{hulse:1975aa} has opened a new window to test the theory of gravitation.  Observations of this pulsar have allowed one to measure the relativistic advance of the periastron \cite{taylor:1976aa} and more importantly to measure the rate of orbital decay due to gravitational radiation \cite{taylor:1979aa}. Pulsars are rotating neutron stars that are emitting very strong radiation. The periods of pulsars are very stable which allows us to consider them as ``clocks''  that are moving in an external gravitational field (typically in the gravitational field generated by a companion). The measurements of the pulse time of arrivals can be used to infer several parameters by fitting an appropriate timing model (see for example Section 6.1 from \cite{will:2014la}): (i) non-orbital parameter such as the pulsar period and its rate of change; (ii) five Keplerian parameters and (iii)  some post-Keplerian parameters \cite{damour:1986fk}. In GR, the expressions of these post-Keplerian parameters are related to the masses of the two bodies and to the Keplerian parameters. If more than 2 of these post-Keplerian parameters can be determined, they can be used to test GR \cite{stairs:2003aa}. Nowadays, more than 70 binary pulsars have been observed \cite{lorimer:2008aa}. A description of the most interesting binary pulsars in order to test the gravitation theory can be found in Section 6.2 from \cite{will:2014la} or in the supplemental material from \cite{shao:2014rc}. 

The model fitted to the observations is based on a post-Newtonian analytical solution to the 2 body equations of motion \cite{damour:1985uq} (see also \cite{wex:1995kq}) and includes contribution from the Einstein time delay (i.e. the transformation between proper and coordinate time), the Shapiro time delay, the Roemer time delay \cite{damour:1986fk}. The model also corrects for several systematics like atmospheric delay, Solar system dispersion, interstellar dispersion, motion of the Earth and the Solar System, \dots (see for example \cite{edwards:2006aa}).

Pulsars observations provide some of the best current constraints on alternative theories of gravitation (for a review, see \cite{wex:2014aa,kramer:2016aa}). In addition to the Hulse and Taylor pulsar, the double pulsar \cite{kramer:2006aa} now provides the best measurement of the pulsar orbital rate of change \cite{kramer:2016aa}.  In addition, the post-Keplerian modeling has been fully derived in tensor-scalar theories \cite{damour:1992ys,damour:1992ve,damour:1996uq} such that pulsars observations have provided some of the best constraints on this class of theory  \cite{freire:2012aa,ransom:2014aa,kramer:2016aa}. It is important to mention that non perturbative strong field effects may arise in binary pulsars system and needs to be taken into account \cite{damour:1993vn,damour:1996uq}.

In addition, binary pulsars have also  been successfully used  to test Lorentz symmetry. For example, analyses of the pulses time of arrivals provide a constraint on the $\alpha_{1,2,3}$ PPN parameters. Since non perturbative strong field effects may arise in binary pulsars system (see for example \cite{foster:2007aa} for strong field effects in Einstein-Aether theory), the obtained constraints are interpreted as strong field version of the PPN parameters denoted by $\hat \alpha_i$. Estimates of these parameters should be compared carefully to the standard weak field constraints since they may depend on the gravitational binding energy of the neutron star. The best current constraint on $\hat \alpha_1=-0.4^{+3.7}_{-3.1}\times 10^{-5}$ is obtained by considering the orbital dynamics of the binary pulsars PSR J1738+0333 \cite{wex:2007aa,shao:2012fk}. The best current constraint on $\hat \alpha_2$ takes advantage from the fact that this parameter produces a precession of the spin axis of massive bodies \cite{nordtvedt:1987cr}. The combination of observations of two solitary pulsars lead to the best current constraints on $\left| \hat \alpha_2\right|<1.6\times 10^{-9}$ \cite{shao:2013aa}. Finally, the parameter $\hat \alpha_3$ produces a violation of the momentum conservation in addition to a violation of the Lorentz symmetry. This parameter will induce a self-acceleration for rotating body that can be constrained using binary pulsars \cite{bell:1996bh}. The best current constraint uses a set of 5 pulsars (4 binary pulsars and one solitary pulsar) and is given by $\hat \alpha_3 < 5.5 \times 10^{-20}$ \cite{gonzalez:2011aa}.

Furthermore, specific Lorentz violating theories have also been constrained with binary pulsars. In \cite{yagi:2014fk,yagi:2014jk}, binary pulsars observations are used to constrain Einstein-Aether and khronometric theory. In these theories, the low-energy limit Lorentz violations can be parametrized by four parameters: the $\alpha_1$ and $\alpha_2$ PPN parameters and two other parameters. It has been shown \cite{foster:2007aa,yagi:2014fk,yagi:2014jk} that the orbital period decay depends on these four parameters. Assuming the solar system constraints on $\alpha_1$ and $\alpha_2$ \cite{will:2014la}, measurements of the rate of change of the orbital period of binary pulsars have been used to constrain the two other parameters (see for example Fig.~2 from \cite{yagi:2014fk}). In this work, strong field effects have been taken into account by solving numerically the field equations in order to determine the neutron stars sensitivity \cite{yagi:2014jk}.

Finally, binary pulsars have been used in order to derive constraints on the SME coefficients. As in the PPN formalism, constraints obtained from binary pulsars need to be considered as constraints on strong-field version of the SME coefficients that may include non perturbative effects. Two different types of effects have been used to determine estimates on the SME coefficients: (i) tests using the spin precession of solitary pulsars and (ii) tests using effects on the orbital dynamics of binary pulsars \cite{shao:2014rc}. The SME contribution to the precession rate of an isolated spinning body has been derived in \cite{bailey:2006uq} and is given by
\begin{equation}
	\Omega^k_\textrm{SME}=\frac{\pi}{P}\bar s^{kj}\hat S^j \, ,
\end{equation}
where $P$ is the spin period and $\hat S^j$ is the unit vector pointing along the spin direction. The effects from the pulsar spin precession on the pulse width can be found in \cite{lorimer:2004aa,shao:2013aa}. Two solitary pulsars have been used to constrain the SME coefficients with this effect. The second type of tests come from the orbital dynamics of binary pulsars. As mentioned in Sections~\ref{sec:llr} and \ref{sec:planets}, the SME will modify the two-body equations of motion by including the term from Eq.~(\ref{eq:eom}). At first order in the SME coefficients, this will produce several secular effects that have been computed in \cite{bailey:2006uq}. In particular, an additional advance in the argument of periastron and of the longitude of the nodes has been mentioned in Eq.~(\ref{eq:SMEad}) and used to constrain the SME with planetary ephemerides. For binary pulsars, it is possible to constrain a secular evolution of two other orbital elements: the eccentricity and the projected semi-major axis $x$. The secular SME contributions to these quantities have been computed in \cite{bailey:2006uq,shao:2014qd,shao:2014rc} and are given by
\begin{subequations}\label{eq:SMEad2}
	\begin{eqnarray}
		\left<\frac{d e}{dt}\right>&=&-n\sqrt{1-e^2}\left[\frac{\varepsilon^2}{e^3}\bs_{PQ}-2\frac{\delta m}{M}\frac{na\varepsilon}{e^2}\bs^P\right]\, ,\\
	\left<\frac{dx}{dt}\right>&=& n \frac{m_C}{m_P+m_C} a \cos i \frac{\varepsilon}{e^2\sqrt{1-e^2}}\left[\bs_{kP}\cos \omega -\sqrt{1-e^2} \bs_{kQ} \sin \omega +2\frac{\delta m}{M}nae\bs^k \cos\omega\right]	\, ,
	\end{eqnarray}
\end{subequations}
where $m_P$ is the mass of the pulsar, $m_C$ is the mass of the companion and all others quantities have been introduced after Eqs.~(\ref{eq:SMEad}). For each binary pulsar, in principle 3 tests can be constructed by using $\dot\omega$, $\dot e$, $\dot x$. In \cite{shao:2014rc}, 13 pulsars have been used to derive estimates on the SME coefficients.  The combination of the observations from the solitary pulsars and from the 13 binary pulsars are reported in Table~\ref{tab:binary}. Both orbital dynamics and spin precession are completely independent of $\bs^{TT}$ whose constraint will be discussed later.

Several comments can be made about this analysis. First of all, it can be considered as a postfit analysis done after an initial fit performed in GR (or within the post-Keplerian formalism). In particular, correlations between the SME coefficients and other parameters (e.g. orbital parameters) are neglected. Secondly, for most of the pulsars, $\dot x$ $\dot \omega$ and $\dot e$ are not directly measured from the pulse time of arrivals but rather estimated from the uncertainties on $x$, $\omega$ and $e$ divided by the time span of the observations. Further, it is important to mention that effects of Lorentz violations have been considered only for the orbital dynamics but never on the Einstein delay or on the Shapiro time delay in this analysis. The full timing model within SME can be found in Section V.E.3 from \cite{bailey:2006uq} (see also \cite{jennings:2015aa} for a similar derivation with the matter-gravity couplings). In addition, some parameters are not measured like for example the longitude of the ascending node $\Omega$ or the azimuthal angle of the spin. These parameters have been marginalized by using Monte Carlo simulations. It is unclear what type of prior probability distribution function has been used in this analysis and what is the impact of this choice. Nevertheless, the results obtained by this analysis (which does not include the $\bar s^{TT}$ parameter) are amongst the best ones currently available demonstrating the power of pulsars observations. The main advantages of using binary pulsars come from the fact that their orbital orientation vary which allows one to disentangle the different SME coefficients and to end up with low correlations. Furthermore, they are so far the only constraints on the strong field version of the SME coefficients.

\begin{table}[H]
\caption{Estimation of SME coefficients from binary pulsars data analysis from \cite{shao:2014qd,shao:2014rc}. No correlations coefficients have been derived in this analysis. These estimates should be considered as estimates on the strong field version of the SME coefficients that may include non perturbative strong field effects due to the gravitational binding energy.}
\label{tab:binary}
\small 
\centering
\begin{tabular}{lr}
\toprule
\textbf{Coefficient}	& \\
\midrule
$\left|\bs^{TT}\right|$        & $< 2.8 \times 10^{-4\phantom{1}}$              \\
$\bs^{XX}-\bs^{YY}$            & $( 0.2 \pm 9.9)     \times 10^{-11}$            \\
$\bs^{XX}+\bs^{YY}-2\bs^{ZZ}$  & $( -0.05 \pm 12.25) \times 10^{-11}$            \\
 $\bs^{XY}$                    & $( 0.05 \pm 3.55)   \times 10^{-11}$            \\
 $\bs^{XZ}$ 	               & $( 0.0 \pm 2.0)     \times 10^{-11}$            \\
 $\bs^{YZ}$ 	               & $( 0.0 \pm 3.3)     \times 10^{-11}$            \\
 $\bs^{TX}$	                   & $( 0.05 \pm 5.25)   \times 10^{-9\phantom{1}}$  \\
 $\bs^{TY}$		               & $( 0.5 \pm 8.0)      \times 10^{-9\phantom{1}}$ \\
 $\bs^{TZ}$ 	               & $( -0.05 \pm 5.85)   \times 10^{-9\phantom{1}}$ \\
\bottomrule
\end{tabular}
\end{table}

In addition,  a different analysis has been performed to constrain the parameter $\bar s^{TT}$ alone \cite{shao:2014qd}. While the orbital dynamics and the spin precession is completely independent of $\bar s^{00}$ (i.e. the time component of $\bs^{\mu\nu}$ in a local frame), the boost between the Solar System and the binary pulsar frame makes appear explicitly the $\bar s^{TT}$ coefficient.  In \cite{shao:2014qd}, the assumption that there exists a preferred frame where the $\bar s^{\mu\nu}$ tensor is isotropic is made, which makes the results specific to that case (although the analysis can be done without this assumption). The analysis requires the knowledge of the pulsar velocity with respect to the preferred frame as well as the velocity of the Solar System with respect to the same frame. Three pulsars have their radial velocity measured, which combined with proper motion in the sky can be used to determine their velocity. The velocity of the Solar System is taken as its velocity with respect to the Cosmic Microwave Background (CMB) frame $\bm w_\odot$ (with $\left|\bm w_\odot\right|=369$ km/s). The analysis is completely similar to the ones performed for the other SME coefficients (see the discussion in the previous paragraph). It is known that $\bs^{TT}$ has a strong effect on the propagation of the light neglected in \cite{shao:2014qd}, which may impact the result. In addition, all correlations between $\bs^{TT}$ and the other SME coefficients are neglected. Finally, two different scenarios have been considered regarding the preferred frame: (i) a scenario where the preferred frame is assumed to be the CMB frame and (ii) a scenario where the orientation of the preferred frame is left free and is marginalized over but the magnitude of the velocity of the Solar System with respect to that frame is still assumed to be the 369km/s. The general case corresponding to a completely free preferred frame has not been considered. If the CMB frame is assumed to be the preferred frame, the constraint on $\bs^{TT}$ is given by $\left|\bs^{TT}\right|<1.6\times 10^{-5}$ which is a bit better than the one obtained with VLBI (see Eq.~(\ref{eq:vlbi})) although the VLBI analysis does not assume any preferred frame. The scenario where the orientation of the preferred frame is left as a free parameter leads to an upper bound on $\left|\bs^{TT}\right|<2.8 \times 10^{-4}$. 

In summary,  observations of binary pulsars are an incredible tool to test the gravitation theory. These tests are of the same order of magnitude (and sometimes better) than the ones performed in the Solar System. Moreover, observations of binary pulsars are sensitive to strong field effects. Observations of the pulse arrival times have been used to search for a breaking of Lorentz violation within the PPN framework by constraining the strong field version of the $\alpha_i$ parameters. The parameter $\hat \alpha_1$ is constrained at the level of $10^{-5}$, $\hat \alpha_2$ at the level of $10^{-9}$ and $\hat \alpha_3$ at the level of $10^{-20}$ \cite{wex:2014aa}. In addition, constraints on Einstein-Aether and khronometric theory have also been done by combining Solar System constraints with binary pulsars observations \cite{yagi:2014fk,yagi:2014jk}. Finally, within the SME framework, a postfit analysis has been done by considering the spin precession of solitary pulsars and the orbital dynamics of binary pulsars.  The obtained results are given in Table~\ref{tab:binary} and constrain the strong field version of the SME coefficients. The main advantage of using binary pulsars comes from the fact that they proved an estimate of all the SME coefficients with reasonable correlations. It has to be noted that the modification of the orbital period due to gravitational waves emission has not been computed so far in the SME formalism. In addition, the constraint on $\bar s^{TT}$ suffers from the assumption of the existence of a preferred frame. Moreover, the corresponding analysis has neglected all effects on the timing delay that may also impact the results and has neglected the other SME coefficients that may also impact this constraint.

\subsection{\v Cerenkov radiation}\label{sec:cerenkov}
Gravitational \v Cerenkov radiation is an effect that occurs when the velocity of a particle exceeds the phase velocity of gravity.  In this case, the particle will emit gravitational radiation until the particle loses enough energy to drop below the gravity speed \cite{kostelecky:2015db}. In modified theory of gravity, the speed of gravity in a vacuum may be different from the speed of light and \v Cerenkov radiation may occur and produces energy losses for particles traveling over long distances. Observations of high energy cosmic rays that have not lost all their energy through \v Cerenkov radiation can be used to put constraints on models of gravitation that predicts gravitational waves that are propagating slower than light. This effect has been used to constrain some alternative gravitation theories~\cite{moore:2001aa,kiyota:2015aa}: a class of tensor-vector theories \cite{elliott:2005aa}, a class of tensor-scalar theories \cite{kimura:2012aa}, extended theories of gravitation~\cite{de-laurentis:2012aa} and some ghost-free bigravity \cite{kimura:2016aa}. 

The propagation of gravitational waves within the SME framework has been derived in \cite{kostelecky:2015db,kostelecky:2016nx} (including nonminimal SME contributions). In particular, in the minimal SME, the dispersion relation for the gravitational waves is given by \cite{kostelecky:2015db}
\begin{equation}\label{eq:speed}
	l_0^2=\left|\bm l\right|^2 +\bs^{\mu\nu}l_\mu l_\nu \, ,
\end{equation}
where $l^\alpha$ is the 4-momentum of the gravitational wave. A similar expression  including nonminimal higher order SME terms can be found in \cite{kostelecky:2015db,kostelecky:2016nx}. If the minimal SME produces dispersion-free propagation, the higher order terms lead to dispersion and birefringence \cite{kostelecky:2016nx}. As can be directly inferred from the last equation, gravitational \v Cerenkov radiation can arise when the effective refractive index $n$ is
\begin{equation}
	n^2=1-\bs^{\mu\nu}\hat l_\mu \hat l_\nu > 1 \, ,
\end{equation}
where $\hat l_\mu=l_\mu/\left|\bm l\right|$. The expression for the energy loss rate due to Lorentz-violating gravitational \v Cerenkov emission has been calculated from tree-level graviton emission for photons, fermions and scalar particles and is given by \cite{kostelecky:2015db}
\begin{equation}\label{eq:dEdt}
	\frac{dE}{dt}=-F^w(d) G \left(\bar s ^{(d)}(\hat {\bm p})\right)^2 \left|\bm p\right|^{2d-4}\, ,
\end{equation}
where $d$ is the dimension of the Lorentz violating operator ($d=4$ for the minimal SME), $F^w(d)$ is a dimensionless factor depending on the flavor $w$ of the particle emitting the radiation, $\bm p$ is the particle incoming momentum (with $\hat {\bm p}=\bm p/\left|\bm p\right|$) and $\bar s^{(d)}$ is a direction-dependent combination of SME coefficients. In the minimal SME, $\bar s^{(4)}(\hat {\bm p})$ is decomposed on spherical harmonics as
\begin{equation}
	\bar s^{(4)}(\hat {\bm p})=\sum_{jm}Y_{jm}(\hat p)\bs^{(SH)}_{jm}\, ,
\end{equation}
where we explicitly indicated the $(SH)$ to specify that these coefficients are spherical harmonic decomposition of the SME coefficients. The calculation of the dimensionless factor $F^w(d)$ for scalar particles, fermions and photons has been done in \cite{kostelecky:2015db}. The integration of Eq.~(\ref{eq:dEdt}) shows that if a cosmic ray of specie $w$ is observed on Earth with an energy $E_f$ after traveling a distance $L$ along the direction $\hat{\bm p}$, this implies the following constraint on the SME coefficients
\begin{equation}\label{eq:cerenkov}
	\bar s^{(d)}(\hat{\bm p})< \sqrt{\frac{\mathcal F^w(d)}{G E_f^{2d-5}L}} \, ,
\end{equation}
where $\mathcal F^w(d)=(2d-5)/F^w(d)$ is another dimensionless factor dependent on the matrix element of the tree-level process for graviton emission.

Using data for the energies and angular positions of 299 observed cosmic rays from different collaborations \cite{takeda:1999aa,bird:1995aa,wada:1980aa,abbasi:2008ab,aab:2015aa,winn:1986aa,abbasi:2014aa,pravdin:2005aa},  \citet{kostelecky:2015db} derived lower and upper constraints on 80 SME coefficients, including the nine coefficients from the minimal SME whose constraints are given by the Table~\ref{tab:cerenkov}. In their analysis, they consider the coefficients from the different dimensions separately and did not fit all of them simultaneously. In addition, in the minimal SME, they did a fit for the $\bar s^{TT}$ parameter alone and another fit for the other 8 coefficients. The number of sources and their directional dependence across the sky allow one to disentangle the SME coefficients and to derive two-sided bounds from the Eq.~(\ref{eq:cerenkov}). The only coefficient that is one sided is $\bar s^{TT}$ because it produces isotropic effects. The bounds are severe for these coefficients, on the order of $10^{-13}$.  However, this analysis assumes that the matter sector coefficients vanish.  Furthermore, several assumptions have been made in order to derive the bounds from Table~\ref{tab:cerenkov}. It is assumed that the cosmic ray primaries are nuclei of atomic weight $N=56$ (iron), that the \v Cerenkov radiation is emitted by one of the fermionic partons in the nucleus that carries 10 \% of the cosmic ray energy and that the travel distance of the cosmic ray is 10 Megaparsec (Mpc) \cite{kostelecky:2015db}. Although only conservative assumptions are used for the astrophysical processes involved in the production of high-energy cosmic rays, the observations rely on the sources on the order of 10 Mpc distant, and thus the analysis is of a different nature than a controlled laboratory or even Solar-System test.

\begin{table}[H]
\caption{Lower and upper limits on the SME coefficients decomposed in spherical harmonics derived from \v Cerenkov radiation \cite{kostelecky:2015db}.}
\label{tab:cerenkov}
\small 
\centering
\begin{tabular}{lrr}
\toprule
\textbf{Coefficient}	& Lower bound & Upper bound \\
\midrule
$\bs^{(SH)}_{00}$        & $-3\times 10^{-14}$  &              \\
$\bs^{(SH)}_{10}$        & $-1\times 10^{-13}$  &  $7\times 10^{-14}$ \\
Re $\bs^{(SH)}_{11}$     & $-8\times 10^{-14}$  &  $8\times 10^{-14}$ \\
Im $\bs^{(SH)}_{11}$     & $-7\times 10^{-14}$  &  $9\times 10^{-14}$ \\
$\bs^{(SH)}_{20}$        & $-7\times 10^{-14}$  &  $1\times 10^{-13}$ \\
Re $\bs^{(SH)}_{21}$     & $-7\times 10^{-14}$  &  $7\times 10^{-14}$ \\
Im $\bs^{(SH)}_{21}$     & $-5\times 10^{-14}$  &  $8\times 10^{-14}$ \\
Re $\bs^{(SH)}_{22}$     & $-6\times 10^{-14}$  &  $8\times 10^{-14}$ \\
Im $\bs^{(SH)}_{22}$     & $-7\times 10^{-14}$  &  $7\times 10^{-14}$ \\
\bottomrule
\end{tabular}
\end{table}

For the sake of completeness and to allow an easy comparison with the estimations of the other standard cartesian $\bar s^{\mu\nu}$ coefficients, the following relations give the links between the spherical harmonic decomposition and the standard cartesian decomposition of the SME coefficients:
\begin{subequations}
\begin{align}
	\bs^{(SH)}_{00}&=\frac{4}{3}\sqrt{4\pi}\, \bs^{TT}\, , \\
	\bs^{(SH)}_{10}&=-\sqrt{\frac{16\pi}{3}}\bs^{TZ}\, ,\qquad  \textrm{Re }\bs^{(SH)}_{11}=\sqrt{\frac{8\pi}{3}}\bs^{TX}\, , \qquad  \textrm{Im }\bs^{(SH)}_{11}=-\sqrt{\frac{8\pi}{3}}\bs^{TY}\, , \\
	\bs^{(SH)}_{20}&=-\sqrt{\frac{4\pi}{5}}\frac{1}{3}\left(\bs^{XX}+\bs^{YY}-2\bs^{ZZ}\right)\, ,\quad  \textrm{Re }\bs^{(SH)}_{21}=-\sqrt{\frac{8\pi}{15}}\bs^{XZ}\, , \quad  \textrm{Im }\bs^{(SH)}_{21}=\sqrt{\frac{8\pi}{15}}\bs^{YZ}\, , \\
	\textrm{Re }\bs^{(SH)}_{22}&=\sqrt{\frac{2\pi}{15}}\left(\bs^{XX}-\bs^{YY}\right)\, , \qquad  \textrm{Im }\bs^{(SH)}_{22}=-2\sqrt{\frac{2\pi}{15}}\bs^{XY}\, .
\end{align}
\end{subequations}
In summary, observations of cosmic rays allow one to derive some stringent boundaries on the SME coefficients. The idea is that if Lorentz symmetry is broken, these high energy cosmic rays would have lost energy by emitting \v Cerenkov radiation that has not been observed. The boundaries on the spherical harmonic decomposition of the SME coefficients are given in the Table~\ref{tab:cerenkov} (in order to compare these boundaries to other constraints, they have been transformed into boundaries on standard cartesian SME coefficients in Table~\ref{tab:all}). For the minimal SME, one can limit the isotropic $\bs^{TT}$ (one sided bound) or the other eight other coefficients in $\bs^{\mu\nu}$, but not all the nine simultaneously. These boundaries are currently the best available in the literature at the exception of $\bar s^{TT}$ whose constraint is only one sided. Nevertheless, several assumptions have been made in this analysis and the observations rely on sources located at very high distances. This analysis is therefore of a different nature than the other ones where more control on the measurements is possible.

\subsection{Summary and combined analysis}\label{sec:summary}
To summarize, several measurements have already  successfully been used to constrain the minimal SME in the gravitational sector (i.e. the $\bs^{\mu\nu}$ coefficients):
\begin{itemize}
	\item Atom interferometry \cite{muller:2008kx,chung:2009uq}.
	\item Lunar Laser Ranging \cite{battat:2007uq,bourgoin:2016yu}.
	\item Planetary ephemerides \cite{iorio:2012zr,hees:2015sf}.
	\item Very Long Baseline Interferometry \cite{le-poncin-lafitte:2016yq}.
	\item Gravity Probe B \cite{bailey:2013kq}.
	\item Pulsars timing \cite{shao:2014qd,shao:2014rc}.
	\item \v Cerenkov radiation \cite{kostelecky:2015db,tasson:2016fk}.
\end{itemize}
A detailed description of all these analyses is provided in the previous subsections and the Table~\ref{tab:all} summarizes the current estimates. It is also interesting to combine all these estimations together to provide the best estimates on the SME coefficients. In order to do this, we perform a large least-square fit including all the results from the Table~\ref{tab:all} \textbf{including the covariance matrices} quoted in the previous subsections. The results from the \v Cerenkov radiation are not included since they rely on a very different type of observations. Two combined fits are presented: one without including the pulsars results and one including the pulsars results. This is due to the fact that pulsars are sensitive to a strong version of the SME coefficients that may include non perturbative strong field effects as described in Section~\ref{sec:pulsars}. If this is the case, then the pulsars results cannot be directly combined with the weak gravitational field estimates on the SME coefficients. If no non perturbative strong field effect arises, then the right column from Table~\ref{tab:combined} presents a combined fit that includes these observations as well. The results from Table~\ref{tab:combined} include all the information currently available in the literature on the $\bs^{\mu\nu}$ (estimations and correlation matrices). It can also be noted that the pulsars results improve significantly the marginalized estimations on $\bs^{TY}$ and $\bs^{TZ}$ by reducing strongly the correlation between these two coefficients.

In addition, several measurements have been used to constrain the non-minimal SME sectors:
\begin{itemize}
	\item Short gravity experiment \cite{shao:2015uq,long:2015kx,shao:2016aa}.
	\item \v Cerenkov radiation \cite{kostelecky:2015db}.
	\item Gravitational waves analysis \cite{kostelecky:2016nx}.
\end{itemize}
A review of these measurements can be found in \cite{tasson:2016fk}.

\begin{sidewaystable}[!htb]
\caption{Summary of all estimations of the $\bs^{\mu\nu}$ coefficients.}
\label{tab:all}
\small 
\centering
\begin{tabular}{lrrrrrr}
\toprule
         & Atomic grav.~\cite{chung:2009uq}                         & LLR~\cite{bourgoin:2016yu}                           & Planetary eph.~\cite{hees:2015sf} & Pulsars~\cite{shao:2014qd,shao:2014rc}  & \multicolumn{2}{c}{\v Cerenkov rad.~\cite{kostelecky:2015db}} \\
&&&&& Lower bound & Upper bound \\
\midrule
 $\bs^{TT}$                      &                                                                                        &            &                                            & $< 2.8 \times 10^{-4\phantom{1}}$     &$-6 \times 10^{-15}<$          \\ 
 $\bs^{XX}-\bs^{YY}$	         &$\left( 4.4 \pm 11 \right)  \times  10^{-9}$   & $\left(0.6 \pm 4.2\right)\times 10^{-11}$            & $(-0.8 \pm 2.0)\times 10^{-10}$            & $( 0.2 \pm 9.9)     \times 10^{-11}$           & $-9\times 10^{-14} <$ & $< 1.2\times 10^{-13} $        \\
 $\bs^{XX}+\bs^{YY}-2~\bs^{ZZ} $ &                                                &                                                      & $(-0.8 \pm 2.7)\times 10^{-10}$            & $( -0.05 \pm 12.25) \times 10^{-11}$           &$-1.9\times 10^{-13} <$ &   $< 1.3\times 10^{-13} $    \\
 $\bs^{XY}$        	             & $\left( 0.2 \pm 3.9\right)  \times  10^{-9}$  & $\left(-5.7 \pm 7.7\right)\times 10^{-12}$           & $(-0.3 \pm 1.1)\times 10^{-10}$            & $( 0.05 \pm 3.55)   \times 10^{-11}$                &$-3.9\times 10^{-14} <$ &   $< 6.2\times 10^{-14} $   \\
 $\bs^{XZ}$ 		             & $\left(-2.6 \pm 4.4\right)  \times  10^{-9}$  & $\left(-2.2 \pm 5.9\right)\times 10^{-12}$           & $(-1.0 \pm 3.5)\times 10^{-11}$            & $( 0.0 \pm 2.0)     \times 10^{-11}$        &$-5.4\times 10^{-14} <$ &   $< 5.4\times 10^{-14} $           \\
 $\bs^{YZ}$ 		             & $\left(-0.3 \pm 4.5\right)  \times  10^{-9}$  &                                                      & $(5.5 \pm 5.2)\times 10^{-12}$             & $( 0.0 \pm 3.3)     \times 10^{-11}$            &$-3.9\times 10^{-14} <$ &   $< 6.2\times 10^{-14} $       \\
 $\bs^{TX}$	                     & $\left(-3.1 \pm 5.1\right)  \times  10^{-5}$  & $\left(-0.9\pm 1.0\right)\times 10^{-8\phantom{0}}$  &  $(-2.9 \pm 8.3)\times 10^{-9\phantom{1}}$ & $( 0.05 \pm 5.25)   \times 10^{-9\phantom{1}}$    &$2.8\times 10^{-14} <$ &   $< 2.8\times 10^{-14} $     \\
 $\bs^{TY}$		                 & $\left( 0.1 \pm 5.4\right)  \times  10^{-5}$  &                                                      & $(0.3 \pm 1.4)\times 10^{-8\phantom{1}}$   & $( 0.5 \pm 8.0)      \times 10^{-9\phantom{1}}$    &$3.1\times 10^{-14} <$ &   $< 2.4\times 10^{-14} $     \\
 $\bs^{TZ}$ 		             & $\left( 1.4 \pm 6.6\right)  \times  10^{-5}$  &                                                      & $(-0.2 \pm 5.0)\times 10^{-8\phantom{1}}$ & $( -0.05 \pm 5.85)   \times 10^{-9\phantom{1}}$    &$1.7\times 10^{-14} <$ &   $< 2.4\times 10^{-14} $     \\
$ \bs^{TY}+0.43~\bs^{TZ}$        &                                               & $\left(6.2 \pm 7.9\right)\times 10^{-9\phantom{0}}$ \\
$ \bs^{XX}+\bs^{YY}-2\bs^{ZZ}-4.5~\bs^{YZ}$ &                                  & $\left(2.3 \pm 4.5\right)\times 10^{-11}$\\
\bottomrule
\end{tabular}
\vspace*{1cm}

\begin{tabular}{lrr}
\toprule
         & VLBI~\cite{le-poncin-lafitte:2016yq} & GPB~\cite{bailey:2013kq}  \\
\midrule
 $\bs^{TT}$                      &   $(-5\pm 8)\times 10^{-5}$ \\
$\bs^{TT}+970\left(\bs^{XX}-\bs^{YY}\right) -0.05 \left(\bs^{XX}+\bs^{YY}-2\bs^{ZZ}\right)+2895\, \bs^{XY} - 3235\, \bs^{XZ}-11\, 240 \,\bs^{YZ} $    &   & $(0.7 \pm 3.1)\times 10^{-3}$ \\ 
$\bs^{XX}-\bs^{YY}+3.02\, \bs^{XY}+0.32\, \bs^{XZ}+1.05 \, \bs^{YZ} $   &   & $(-1.1 \pm 3.8)\times 10^{-7}$\\ 
\bottomrule
\end{tabular}
\end{sidewaystable}

\begin{table}[htb]
\caption{Estimation of SME coefficients resulting from a fit combining results from: atomic gravimetry (see Table~\ref{tab:AI}), VLBI (see Eq.~\ref{eq:vlbi}), LLR (see Table~\ref{tab:bourgoin}), planetary ephemerides (see Table~\ref{tab:planets}), Gravity Probe B (see Table~\ref{tab:GPB}). The correlation matrices from all these analyses have been used in the combined fit. The right column includes the pulsars results from Table~\ref{tab:binary} as well. The three estimates on $\bs^{JJ}$ are obtained by using the traceless condition $\bs^{TT}=\bs^{XX}+\bs^{YY}+\bs^{ZZ}$.}
\label{tab:combined}
\small 
\centering
\begin{tabular}{lrr}
\toprule
\textbf{Coefficient}	& \textbf{Without pulsars} & \textbf{With pulsars} \\
\midrule
$\bs^{TT}$                     & $( -5. \pm 8.)     \times 10^{-5\phantom{1}}$     & $( -4.6 \pm 7.7)     \times 10^{-5\phantom{1}}$  \\
$\bs^{XX}-\bs^{YY}$            & $( -0.5 \pm 1.9)     \times 10^{-11}$             & $( -0.5 \pm 1.9)     \times 10^{-11}$           \\
$\bs^{XX}+\bs^{YY}-2\bs^{ZZ}$  & $( 1.6 \pm 3.1)      \times 10^{-11}$             & $( 0.8 \pm 2.5)      \times 10^{-11}$           \\
 $\bs^{XY}$                    & $( -1.5 \pm 6.8)     \times 10^{-12}$             & $( -1.6 \pm 6.6)     \times 10^{-12}$           \\
 $\bs^{XZ}$ 	               & $( -1.0 \pm 4.1)     \times 10^{-12}$             & $( -0.8 \pm 3.9)     \times 10^{-12}$           \\
 $\bs^{YZ}$ 	               & $( 2.6 \pm 4.7)     \times 10^{-12}$              & $( 1.1 \pm 3.2)     \times 10^{-12}$           \\
 $\bs^{TX}$	                   & $( -0.1 \pm 1.3)   \times 10^{-9\phantom{1}}$     & $( -0.1 \pm 1.3)   \times 10^{-9\phantom{1}}$  \\
 $\bs^{TY}$		               & $( 0.5 \pm 1.1)      \times 10^{-8\phantom{1}}$   & $( 0.4 \pm 2.3)      \times 10^{-9\phantom{1}}$\\
 $\bs^{TZ}$ 	               & $( -1.2 \pm 2.7)   \times 10^{-8\phantom{1}}$     & $( -0.6 \pm 5.5)   \times 10^{-9\phantom{1}}$  \\
\midrule                                                                          
$\bs^{XX}$                    &  $( -1.7 \pm 2.7)     \times 10^{-5\phantom{1}}$     & $( -1.5 \pm 2.6)     \times 10^{-5\phantom{1}}$  \\
$\bs^{YY}$                    &  $( -1.7 \pm 2.7)     \times 10^{-5\phantom{1}}$     & $( -1.5 \pm 2.6)     \times 10^{-5\phantom{1}}$  \\
$\bs^{ZZ}$                    &  $( -1.7 \pm 2.7)     \times 10^{-5\phantom{1}}$     & $( -1.5 \pm 2.6)     \times 10^{-5\phantom{1}}$  \\
\bottomrule
\end{tabular}
\end{table}

\FloatBarrier
\section{The future}\label{sec:future}
In addition to all the improvements related to existing analysis suggested in the previous sections, there are a couple of sensitivity analyses that have been done within the SME framework. First of all, a thorough and detailed analysis of a lot of observables related to gravitation can be found in \cite{bailey:2006uq}. In addition, we will present in the next subsections a couple of analyses and ideas that may improve the SME coefficients estimates in the future.

\subsection{The Gaia mission}\label{sec:gaia}
 Launched in December 2013, the ESA Gaia mission \cite{de-bruijne:2012kx} is scanning regularly the whole celestial sphere once every 6 months providing high precision astrometric data for a huge number ($\approx$ 1 billion) of celestial bodies. In addition to stars, it is also observing Solar System Objects (SSO), in particular asteroids. The high precision astrometry (at sub-mas level) will allow us to perform competitive tests of gravitation and to provide new constraints on alternative theories of gravitation.

First of all, the Gaia mission is expected to provide an estimate of the $\gamma$ PPN parameter at the level of $10^{-6}$~\cite{mignard:2010kx} by measuring the deflection of the light on a 5 years timescale. Furthermore, in addition to this global determination of a global PPN parameter from observations of light deflection, it has been proposed to use Gaia observations to map the deflection angle in the sky and to look for a dependence of the $\gamma$ PPN parameter with respect to the Sun impact parameter \cite{jaekel:2005vn,jaekel:2005zr,jaekel:2006kx,reynaud:2007fk,reynaud:2009fk}. Such a dependence of the gravitational deflection with respect to the observation geometry is also a feature predicted by SME as shown in \cite{tso:2011uq}. Therefore, the global mapping of the light deflection with Gaia can also be efficiently used to constrain some SME coefficients. A first sensitivity analysis can be found in \cite{tso:2011uq} and is reported on Table~\ref{table:deflection}. Note that proposals observations and missions like AGP \cite{gai:2012fk} or LATOR \cite{turyshev:2007it} can in the long term improve these estimates further by improving the light deflection measurement.

\begin{table}[htb]
\caption{Sensitivity of the SME coefficients to the measurement of the light deflection by several space missions or proposals (these estimates are based on Table I from \cite{tso:2011uq}).}
\label{table:deflection}
\small 
\centering
\begin{tabular}{lccc}
\toprule
\textbf{Mission}	& $\bs^{TT}$ & $\bs^{TJ}$ & $\bs^{IJ}$ \\
\midrule
Gaia \cite{de-bruijne:2012kx} & $10^{-6}$  &  $10^{-6}$ & $10^{-5}$ \\
AGP \cite{gai:2012fk}         & $10^{-7}$  &  $10^{-7}$ & $10^{-6}$ \\
LATOR \cite{turyshev:2007it}  & $10^{-8}$  &  $10^{-8}$ & $10^{-7}$ \\
\bottomrule
\end{tabular}
\end{table}

\begin{figure}[H]
\centering
\includegraphics[width=0.3\textwidth]{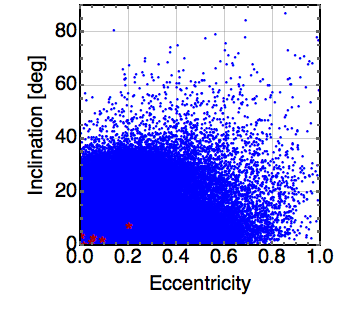}\hspace{0.15\textwidth} \includegraphics[width=0.3\textwidth]{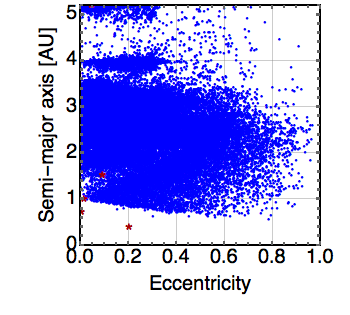} 
\caption{This figure represents the distribution of the orbital parameters for the SSOs expected to be observed by the Gaia satellite. The red stars represent the innermost planets of the Solar System.}
\label{fig:aster}
\end{figure}

In addition to gravitation tests performed by measuring the light deflection, Gaia also provides a unique opportunity to test gravitation by considering the orbital dynamics of SSO.  One can estimate that about 360 000 asteroids will be regularly observed by Gaia at the sub-mas level, which will allow us to perform various valuable tests of gravitation~\cite{mouret:2011uq,hees:2015rc}. In particular, realistic simulations of more than 250 000 asteroids have shown that Gaia will be able to constrain the $\beta$ PPN parameter at the level of $10^{-3}$~\cite{mouret:2011uq}. The main advantage from Gaia is related to the huge number of bodies that will be observed with very different orbital parameters as illustrated on Figure~\ref{fig:aster}. As a consequence, the huge correlations appearing in the planetary ephemerides analysis (see Section~\ref{sec:planets}) will not appear when considering asteroids observations and the marginalized confidence intervals will be highly improved compared to planetary ephemerides analysis.

A realistic sensitivity analysis of Gaia SSOs observations within the SME framework has been performed (see also \cite{hees:2015rc} for preliminary results). In this analysis, 360 000 asteroids have been considered over the nominal mission duration (i.e. five years) and a match between the SSO trajectories with the Gaia scanning law is performed to find the observation times for each SSO. Simultaneously with the equations of motion, we integrate the variational equations, the simulated SSO trajectories being transformed into astrometric observables as well as their partial derivatives with respect to the parameters considered in the covariance analysis. The covariance analysis leads to the estimated uncertainties presented in Table~\ref{tab:gaia}. These uncertainties are incredibly good, which is due to the variety of the asteroids orbital parameters as discussed above. Using our set of asteroids, the correlation matrix for the SME parameters is very reasonable: the most important correlation coefficients are 0.71, -0.68 and 0.46. All the other correlations are below 0.3. Therefore, Gaia offers a unique opportunity to constrain Lorentz violation through the SME formalism.  Finally, the Gaia mission is likely to be extended to 10 years, therefore doubling the measurements baseline which will also impact significantly the expected uncertainties. Finally, it is worth mentioning that the Gaia dataset can be combined with radar observations \cite{margot:2010fk} that are complementary in the time frame and orthogonal to astrometric telescopic observations.

\begin{table}[htb]
\caption{Sensitivity of the SME coefficients to the observations of 360 000 asteroids by the Gaia satellite during a period of 5 years.}
\label{tab:gaia}
\small 
\centering
\begin{tabular}{lccc}
\toprule
\textbf{SME coefficients} & \textbf{Sensitivity} ($1-\sigma$)\\
\midrule
  $\bar s^{XX}-\bar s^{YY}$                 &   $3.7 \times 10^{-12}$             \\
  $\bar s^{XX}+\bar s^{YY}-2\bar s^{ZZ}$    &   $6.4 \times 10^{-12}$             \\
  $\bar s^{XY}$                             &   $1.6 \times 10^{-12}$             \\
  $\bar s^{XZ}$                             &   $9.2 \times 10^{-13}$             \\
  $\bar s^{YZ}$                             &   $1.7 \times 10^{-12}$             \\
  $\bar s^{TX}$                             &   $5.6 \times 10^{-9\phantom{1}}$   \\  
  $\bar s^{TY}$                             &   $8.8 \times 10^{-9\phantom{1}}$   \\
  $\bar s^{TZ}$                             &   $1.6 \times 10^{-8\phantom{1}}$   \\
\bottomrule
\end{tabular}
\end{table}

In summary, the Gaia space mission offers two opportunities to test Lorentz symmetry in the Solar System by looking at the deflection of light and by considering the orbital dynamics of SSO. The second type of observations is extremely interesting in the sense that the high number and the variety of orbital parameters of the observed SSO leads to decorrelate the SME coefficients. 

\subsection{Analysis of Cassini conjunction data}
The  space mission Cassini is exploring the Saturnian system since July 2004. During its cruising phase while the spacecraft was on its interplanetary journey between Jupiter and Saturn, a measurement of the gravitational time delay was performed~\cite{bertotti:2003uq}. This measurement occurred during a Solar conjunction in June 2002 and was made possible thanks to a multi-frequency radioscience link (at X and Ka-band) which allows a cancellation of the solar plasma noise~\cite{bertotti:2003uq}. The related data spans over 30 days and has been analyzed in the PPN framework leading to the best estimation of the $\gamma$ PPN parameter so far given by $(2.1\pm 2.3)\times 10^{-5}$~\cite{bertotti:2003uq}.

The exact same set of data can be reduced within the SME framework and is expected to improve our current $\bar s^{TT}$ estimation. The time delay within the SME framework has been derived in \cite{bailey:2009fk} and is given by Eq.~(\ref{eq:delay}). 

\begin{figure}[H]
\centering
 \includegraphics[width=0.6\textwidth]{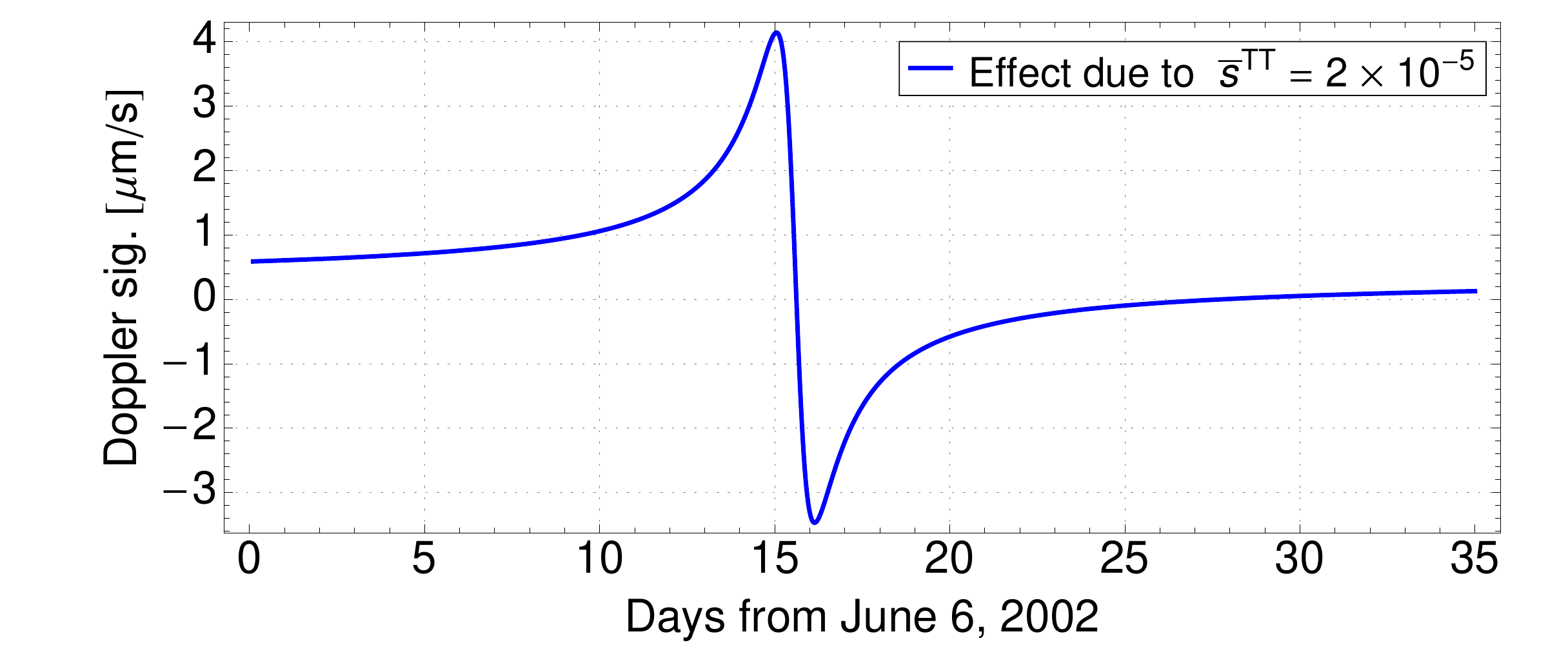} 
\caption{Doppler signature produced by $\bs^{TT}=2 \times 10^{-5}$ on the 2-way Doppler link Earth-Cassini-Earth during the 2002 Solar conjunction.}
\label{fig:cassini}
\end{figure}

A simulation of the Cassini link during the 2002 conjunction within the full SME framework has been realized using the software presented in \cite{hees:2012fk} (see also \cite{hees:2014ys,hees:2015zr}). The signature produced by the $\bar s^{TT}$ coefficients on the 2-way Doppler link during the Solar conjunction is illustrated on Figure~\ref{fig:cassini}. In~\cite{bailey:2009fk}, a crude estimate of attainable sensitivities in estimate of the SME coefficients using the Cassini conjunction data is given (see Table I from \cite{bailey:2009fk}). It is shown that some combinations of the $\bs^{IJ}$ coefficients can only be constrained at the level of $10^{-4}$, which is 7 to 8 orders of magnitude worse than the current best constraints on these coefficients. It is therefore safe to neglect these and to concentrate only on the $\bs^{TT}$ coefficient.  A realistic covariance analysis performed over the 30 days of the Solar conjunction and assuming an uncertainty of the Cassini Doppler of 3 $\mu m/s$ \cite{iess:2007ve,kliore:2004zr} shows that the $\bs^{TT}$ parameter can be constrained at the level of $2\times 10^{-5}$ using the Cassini data allowing an improvement of a factor 4 with respect to the current best estimate coming from VLBI analysis (see Eq.~(\ref{eq:vlbi})). Therefore, a reanalysis of the 2002 Cassini data within the SME framework would be highly valuable.

\subsection{Satellite Laser Ranging (LAGEOS/LARES)}\label{sec:slr}
Searching for violations of Lorentz symmetry by using the orbital motion of planets (see Section~\ref{sec:planets}), binary pulsars (see Section~\ref{sec:pulsars}),  the Moon (see Section~\ref{sec:llr}) and  asteroids (see Section~\ref{sec:gaia}) has turned out to be highly powerful. It is therefore logical to consider the motion of artificial satellite orbiting around Earth to search for Lorentz violations. In particular, laser ranging to the two LAGEOS and  to the LARES satellites has successfully been used to test GR by measuring the impact of the Schwarzschild precession on the motion of the satellites \cite{iorio:2002it,lucchesi:2010pi,lucchesi:2014ph}. It has also been claimed that the impact of the frame-dragging (or Lense-Thirring effect) due to the Earth's spin on the orbital motion of the satellites has been measured \cite{ciufolini:2004uq,ciufolini:2016pi,ciufolini:2009bs,ciufolini:2012sf,ciufolini:2012kx,paolozzi:2011ad} although this claim remains controversial \cite{iorio:2009bf,iorio:2009oa,iorio:2009kl,iorio:2011ys,renzetti:2012qa,renzetti:2013mi}. Similarly, the LAGEOS/LARES satellites can also be used to search for Lorentz violations. A sensitivity analysis has been done in \cite{iorio:2012zr} and it has been shown that the LAGEOS satellites are sensitive at the level of $10^{-4}$ to the $\bar s^{TJ}$ coefficients. Using LARES should improve significantly this value. Further numerical simulations are required in order to determine exactly the SME linear combinations to which  the ranging to these satellites is sensitive to. A data analysis within the full SME framework (i.e. including the integration of the SME equation of motion and including the SME coefficients with the other global parameters in the fit) would also be highly interesting. In addition, similar tests of Lorentz symmetry can also be included within the scientific goals of the  LAser RAnged Satellites Experiment (LARASE) project \cite{lucchesi:2015fk} or within the OPTIS project \cite{lammerzahl:2004uq}.

\subsection{Gravity-Matter coefficients and breaking of the Einstein equivalence principle}\label{sec:gravmat}
All the measurements mentioned in Section~\ref{sec:data} can be analyzed by considering the gravity-matter coupling coefficients $\bar a_\mu$ and $\bar c_{\mu\nu}$ \cite{kostelecky:2011kx} that are breaking the EEP. Some atomic clocks measurements have already provided some constraints on the $\bar a_\mu$ coefficients \cite{hohensee:2011fk,hohensee:2013yg,hohensee:2013fp}. In addition, in \cite{hees:2015sf} the planetary ephemerides analysis is interpreted by considering the $\bar a_\mu$ coefficients and the atomic interferometry results from \cite{chung:2009uq} and the LLR results from \cite{battat:2007uq} are also reinterpreted by considering the gravity-coupling coefficients. Clearly this is a preliminary analysis that needs to be refined by more solid data reductions. Considering the increasing number of fitted parameters, it is of prime importance to increase the number of measurements used in the analysis and to produce combined analysis with as many types of observations as possible. The measurements developed in Section~\ref{sec:data} are a first step in order to reach this goal. The gravity-coupling coefficients can also be constrained by more specific tests related to the EEP like for example tests of the Universality of Free Fall with MicroSCOPE \cite{touboul:2001kx,touboul:2012cr}, tests of the gravitational redshift with GNSS satellites \cite{delva:2015fk}, with the Atomic Clocks Ensemble in Space (ACES) project \cite{cacciapuoti:2011ve}, or with the OPTIS project \cite{lammerzahl:2004uq}, \dots 

\section{Conclusions}\label{sec:conclusion}
Lorentz symmetry is at the heart of both GR and the Standard Model of particle physics. This symmetry is broken in various scenarios of unification, of quantum gravity and even in some models of Dark Matter and Dark Energy. Searching for violations of Lorentz symmetry is therefore a powerful tool to test fundamental physics. The last decades have seen the number of tests of Lorentz invariance arise dramatically in all sectors of physics~\cite{kostelecky:2011ly}. In this review, we focused on searches for Lorentz symmetry breaking in the pure gravitational sector. Mainly two frameworks exist to parametrize violations of Lorentz invariance in the gravitation sector. First of all, the three $\alpha_{1,2,3}$ PPN parameters phenomenologically encode a violation of Lorentz symmetry at the level of the spacetime metric~\cite{will:2014la}. These parameters are constrained by LLR (see Section~\ref{sec:llr}) and by pulsars timing measurements (see Section~\ref{sec:pulsars}). In addition, it is interesting to notice that the corresponding PPN metric parametrizes also Einstein-Aether and Khronometric theories in the weak gravitational field limit \cite{yagi:2014fk} while these theories have a more complex strong field limit (and can show non perturbative effects) that have been constrained by pulsars observations (see Section~\ref{sec:pulsars} and \cite{yagi:2014fk,yagi:2014jk}).

In addition to the PPN formalism, the SME formalism has been developed by including systematically all possible Lorentz violations terms that can be constructed at the level of the action. In the pure gravitational sector, the gravitational action within the SME formalism contains the usual Einstein-Hilbert action but also new Lorentz violating terms constructed by contracting new fields with some operators built from curvature tensors and covariant derivatives with increasing mass dimension~\cite{bailey:2016aa}. The lower mass dimension term is known as the minimal SME. In the limit of linearized gravity, the observations within the minimal SME formalism depend on 9 coefficients, the $\bs^{\mu\nu}$ symmetric traceless tensor. This formalism offers a new opportunity to search for deviations from GR in a framework different from the standard PPN formalism. We reviewed the different observations that have been used so far to constrain the SME coefficients. The main idea is to search for a signature (usually periodic) that arises from a dependence on the orientation of the system measured (the dependence on the orientation is typically due to the Earth's rotation, the orbital motion of the planets around the Sun, etc\dots) or from a dependence on the boost of the system observed (so far, only the binary pulsars $\bs^{TT}$ constraint comes from this type of dependence \cite{shao:2014qd}). Most of SME analyses are postfit analyses in the sense that analytical signatures due to SME are fitted in residuals noise obtained in a previous data reduction performed in pure GR. In Section~\ref{sec:postfit}, we showed that this approach can sometimes lead to overoptimistic constraint on the SME coefficients and that one should be careful in interpreting results obtained using such an approach.

In Section~\ref{sec:data}, we discussed in details the different measurements used so far to constrain the $\bs^{\mu\nu}$ coefficients: atomic gravimetry (Section~\ref{sec:ai}), VLBI (Section~\ref{sec:vlbi}), LLR (Section~\ref{sec:llr}), planetary ephemerides (Section~\ref{sec:planets}), Gravity Probe B (Section~\ref{sec:gpb}), pulsars timing (Section~\ref{sec:pulsars}) and \v Cerenkov radiation (Section~\ref{sec:cerenkov}). In each of these subsections, we describe the  current analyses performed in order to constrain the SME coefficients and provide a critical discussion from each of them. We also provide a summary of these constraints on Table~\ref{tab:all}. In addition, we used all these results to produce a combined analysis of the SME coefficients. This fit is done by taking into account the correlation matrices for each individual analysis. The results of this combined fit are presented in Table~\ref{tab:combined} and are the current best estimates of the SME coefficients that are possible to derive with all available analyses. In addition to the minimal SME, there exists higher order Lorentz violating terms that have been considered and constrained by short-range gravity experiments \cite{shao:2015uq,long:2015kx,shao:2016aa}, gravitational waves analysis \cite{kostelecky:2016nx} and \v Cerenkov radiation \cite{kostelecky:2015db,tasson:2016fk}.

In Section~\ref{sec:future}, we discussed some opportunities to improve the current constraints on the SME coefficients. In particular, the European space mission Gaia offers an excellent opportunity to probe Lorentz symmetry through the measurement of light deflection and through the orbital motion of asteroids. The Cassini conjunction data also offers a  way to constrain the $\bs^{TT}$ coefficient that impacts severely the propagation of light. Finally, existing satellite laser ranging data can also be analyzed within the SME framework.

In addition, as mentioned in Section~\ref{sec:gravmat}, all the analyses presented in this review can include gravity-matter coefficients~\cite{kostelecky:2011kx}. While considering these, the number of coefficients fitted increase significantly and it becomes crucial to produce a fit combining several kinds of experiments. A preliminary analysis considering these coefficients for planetary ephemerides, LLR and atomic gravimetry has been performed in \cite{hees:2015sf} but needs to be refined. In addition, some atomic clocks experiments have already been used to constrain matter-gravity coefficients \cite{hohensee:2011fk,hohensee:2013yg,hohensee:2013fp}.

In conclusion, though no violation of Lorentz symmetry has been observed so far, an incredible number of opportunities still exists for additional investigations. There remains a large area of unexplored coefficients space that can be explored by improved measurements or by new projects aiming at searching for breaking of Lorentz symmetry. In addition, the increasing number of parameters fitted (by including the gravity-matter coupling coefficients simultaneously with the pure gravity coefficients in the analyses) will deter the marginalized estimates of each coefficient. This verdict emphasizes the need to increase the types of measurements that can be combined together to explore the vast parameters space as efficiently as possible. The current theoretical questions related to the quest for a unifying theory or for a quantum theory of gravitation suggests that Lorentz symmetry will play an important role in the search for new physics. Hopefully, future searches for Lorentz symmetry breaking will help theoreticians to unveil some of the mysteries about Planck-scale physics \cite{tasson:2014qv}.

\vspace{6pt}

\acknowledgments{A.H. is thankful to P. Wolf, S. Lambert, B. Lamine, A. Rivoldini, F. Meynadier, S. Bouquillon, G. Francou, M.-C. Angonin, D. Hestroffer, P. David and A. Fienga for interesting discussions about some part of this work. Q. G. B. was supported in part by the National Science Foundation under Grant No. PHY-1402890. A. B. and C.L.P.L. are grateful for the CNRS/GRAM and ``Axe Gphys'' of Paris Observatory Scientific Council. C. G. and Q. G. B. acknowledge support from Sorbonne Universit\'es Emergence grant.}


\conflictofinterests{The authors declare no conflict of interest.} 

\abbreviations{The following abbreviations are used in this manuscript:\\

\noindent CMB: Cosmic Microwave Background \\
\noindent ELPN: \'Eph\'em\'eride Lunaire Parisienne Num\'erique\\
\noindent GPB: Gravity Probe B\\
\noindent GR: General Relativity\\
\noindent GRAIL: Gravity Recovery And Interior Laboratory \\
\noindent INPOP: Int\'egrateur Num\'erique Plan\'etaire de l'Observatoire de Paris \\
\noindent IVS: International VLBI Service for Geodesy and Astrometry\\
\noindent LARASE:  LAser RAnged Satellites Experiment  \\
\noindent LLR: Lunar Laser Ranging\\
\noindent LIV: Lorentz Invariance Violation \\
\noindent mas: milliarcsecond\\
\noindent Mpc: Megaparsec \\
\noindent NFT: Numerical Fourier Transform  \\
\noindent PPN: Parametrized Post-Newtonian\\
\noindent SME: Standard-Model Extension\\
\noindent SSO: Solar System Object\\
\noindent TAI: International Atomic Time\\
\noindent TDB: Barycentric Dynamical Time\\
\noindent TT: Terrestrial Time\\
\noindent UTC: Universal Time Coordinate\\
\noindent VLBI: Very Long Baseline Interferometry\\
\noindent yr: year
}



\bibliographystyle{mdpi}

\bibliography{SMEreview}

\end{document}